\documentclass[journal]{IEEEtran}
\pdfoutput=1
\usepackage{graphicx}
\usepackage{epstopdf}

\usepackage{threeparttable}
\input epsf
\usepackage{amsmath}
\usepackage{subfigure}
\usepackage{cite} 
\usepackage{times}
\usepackage{url}
\usepackage{amsthm}  
\usepackage{color}
\usepackage{amssymb}

\newtheorem{thm}{Theorem}
\newtheorem{lemma}{Lemma}

\newtheorem{property}{Property}
\newtheorem{definition}{Definition}

\newtheorem{conj}{Conjecture}

\usepackage{setspace}
\usepackage{url}

\begin{document}
\title{Distributed Scheduling Algorithms for High-Speed Switching Systems}
\author{$\dagger$Shunyuan Ye, $\ddagger$Yanming Shen, $\dagger$Shivendra Panwar\\
$\dagger$Department of Electrical and Computer Engineering, Polytechnic Institute of NYU\\
$\ddagger$School of Computer Science and Technology, Dalian University of Technology, China\\
e-mail: sye02@students.poly.edu, shen@dlut.edu.cn, panwar@catt.poly.edu
}

\maketitle
\newcommand{\be}{\begin{itemize}} \newcommand{\ee}{\end{itemize}}
\newcommand{\tb}{\textbf} \newcommand{\ttt}{\texttt}
\newcommand{\tit}{\textit} \newcommand{\uline}{\underline}

\thispagestyle{empty}
\begin{abstract}
Given the rapid increase in traffic, greater demands have been put on research in high-speed switching systems. Such systems have to simultaneously meet several constraints, e.g., high throughput, low delay and low complexity. This makes it challenging to design an efficient scheduling algorithm, and has consequently drawn considerable research interest. However, previous results either cannot provide a $100\%$ throughput guarantee without a speedup, or require a complex centralized scheduler. In this paper, we design a {\it distributed} $100 \%$ throughput algorithm for crosspoint buffered switches, called DISQUO, with very limited message passing. We prove that DISQUO can achieve $100\%$ throughput for any admissible Bernoulli traffic, with a low time complexity of $O(1)$ per port and a few bits message exchanging in every time slot. To the best of our knowledge, it is the first distributed algorithm that can provide a $100\%$ throughput for a crosspoint buffered switch.
\end{abstract}

\section{Introduction}
\label{sec:intro}
With the growing Internet traffic demand, there is an increasing interest in designing large-scale high-performance packet switches. There is also a growing need for high-speed switching in the backplane of multiprocessing high-performance computer architectures \cite{hem,mink}, and in large data centers \cite{fares,helios}. 

A scheduling algorithm is needed to schedule packet transmissions in such a system. A good algorithm has to meet several requirements, e.g., high throughput, low delay, and low complexity. In order to achieve these requirements, such switches usually require centralized, sometimes complex, algorithms. The well-known \emph{maximum weight matching (MWM)} algorithm \cite{mwm01, mwm02} can achieve $100\%$ throughput for any admissible arrival traffic, but it is not practical to implement due to its high computational complexity ($O(N^3)$). Also, the MWM algorithm needs a centralized scheduler. This increases the implementation complexity and leads to communication overhead. A number of practical iterative algorithms have been proposed, such as iSLIP \cite{islip} and DRRM \cite{drrm}. However, they cannot guarantee $100\%$ throughput for general arrival traffic patterns.

Due to the memory speed limit, most current switches use \emph{input queuing (IQ)} \cite{mwm02, islip, pim, maximal} or \emph{combined input and output queuing (CIOQ)} \cite{cioq}. To address the high complexity of designing scheduling algorithms for input-queued switching architectures, the \emph{crosspoint buffered switching} architecture has been proposed, which promises a simpler scheduling algorithms and better delay performance \cite{rr_rr, lqf_rr, sbf}.
For a crosspoint buffered switch, each input maintains virtual output queues (VOQs), one for each output, and each crosspoint contains a finite buffer. With a speedup of 2, the authors in \cite{chuang05, turner06} showed that a crosspoint buffered switch can provide $100\%$ throughput under any admissible traffic. However, without speedup, previous $100\%$ throughput results are only limited to uniform traffic loads. Under uniform traffic, it has been shown that longest queue first at the input port and round-robin at the output port (LQF-RR) \cite{lqf_rr}, or a simple round-robin scheduler at both input and output ports (RR-RR) \cite{rr_rr}, guaranteed 100\% throughput. In \cite{shah05}, the authors proposed a distributed scheduling algorithm and derived a relationship between throughput and the size of crosspoint buffers. However, to achieve $100\%$ throughput, an infinite-size crosspoint buffer was needed. To our knowledge, there is no distributed algorithm that can achieve $100\%$ throughput for a \emph{finite} crosspoint buffer without speedup.

There has always been a close relationship between the field of switch scheduling and scheduling transmissions in wireless ad hoc networks \cite{mwm01, auction}. Recently, it has been shown that CSMA-like algorithms \cite{aloha08, jiang08, ni0908, liu10, ni1008, ghaderi} can achieve the maximum throughput in wireless ad hoc networks. Stations only need to sense the channel and make their scheduling decisions based on local queue information. These algorithms are easy to implement since no message passing is required. They are the first \emph{distributed} algorithms that can achieve the maximum theoretical capacity in wireless networks.

Inspired by the CSMA-like algorithms and using many of the technique pioneered by them, we propose a distributed algorithm for crosspoint buffered switches 
that can stabilize the system under any admissible Bernoulli traffic.  
Note that for such CSMA-like algorithms to work properly, a node has to know its neighborsÕ state in the previous slot by carrier sensing. This can be achieved in a wireless network due to the broadcast property of the medium. 
However, this cannot be easily implemented in a switching system. 
We make a key observation that the crosspoint buffers can be used for implicit message passing. By observing the buffer states, an input can get some partial information on whether the corresponding outputs are busy or not. This requires no change in the switch fabric architecture or implementation. 
Based on this observation, we designed DISQUO; the
DIStributed QUeue input-Output scheduler. With DISQUO, an input only uses its
local queue information and the locally observable partial schedule in the previous time slot
to make its scheduling decision. We prove the stability of
the system and evaluate the performance of DISQUO by
running extensive simulations. For technical reasons, each input does need to have the global maximum queue size in the system, which requires some message exchanging in each time slot. However, the simulations we conducted show that without the explicit message passing, the algorithm can still stabilize the system for the traffic patterns that we tested. Therefore, we propose the fully distributed algorithm without the global maximum queue size information as a conjecture. This result fulfills the long sought conjectured promise of this architecture \cite{chuang05, turner06, lqf_rr, rr_rr, shah05}. The simulation results also show that it can
provide good delay performance, comparable to output-queued
switches, under different types of traffic.

The rest of the paper is organized as follows. Some theoretical preliminaries are presented in the next section. We present DISQUO in Section \ref{sec:disquo} and prove the system stability. Simulation results are presented in Section \ref{sec:simulation}. We give the proof of system convergence in Section \ref{sec:appb}, and conclude the paper in Section \ref{sec:conclusion}.


\section{Preliminaries}
\label{sec:prelim}
In this section, we introduce the notation and preliminary results that we will use in the theoretical proof of our algorithms. 

\subsection{Glauber Dynamics}
A sequence of random variables ($X_0$, $X_1$, $\cdots$) is a \emph{Markov chain} with state space ${\bf \Omega}$ and transition matrix ${\bf P}$ if for all $x$, $y \in {\bf \Omega}$, all $n \geq 1$, and all events $\textrm{
$\mathcal{H}$}_{n-1}$$ = \cup_{s=0}^{n-1}{\{X_s = x_s\}}$, we have 
\begin{eqnarray}
&&P\{X_{n+1} = y | \{X_n = x\} \cup \textrm{$\mathcal{H}$}_{n-1} \}  {} \nonumber \\
{} && = P\{X_{n+1} = y | X_n = x \} = P(x,y)  \nonumber
\end{eqnarray}
The process can then be described as:
\begin{displaymath}
 \textrm{{\boldmath $\mu$}}(\tau) = \textrm{\boldmath $\mu$}(\tau -1) {\bf P} = \textrm{\boldmath $\mu$}(0) {\bf P}^{\tau},
\end{displaymath}
where {\boldmath $\mu$}$(\tau)$ is the probability distribution of $X_{\tau}$. 

The Markov chain is \emph{irreducible} if any state can reach any other state. If the system starts from any state $X$ and it can return to the state within finite time with a probability $1$, the Markov chain is \emph{positive recurrent}. If the Markov chain is irreducible and positive recurrent, it has a unique stationary distribution {\boldmath $\pi$} so that:
\begin{displaymath}
\lim_{ \tau \to \infty} \textrm{\boldmath $\mu$} (\tau) = \textrm{\boldmath $\pi$}.
\end{displaymath}
 
Let $P^*$ denote the transition probability matrix for the reverse Markov chain ($\cdots, X_n, X_{n-1}, \cdots$). If $P = P^*$, the Markov chain is called \emph{time-reversible} \cite{dynamic}.

A \emph{graph} $G = (V, E)$ consists of a \emph{vertex set} $V$ and an \emph{edge set} $E$, where the elements of $E$ are unordered pair of vertices: $E \subset \{ \{x, y\}: x, y \in V, x \neq y \}$. If $\{x, y\} \in E$, $y$ is a \emph{neighbor} of $x$ (and also $x$ is a neighbor of $y$).  Let $\mathcal{N}(x)$ denote all the neighbors of $x$. A \emph{independent set} $I \subset V$ is a set such that if $x \in I$, $\forall y \in \mathcal{N}(x)$, $y \notin I$. Let $\mathcal{I}(G)$ represent all the independent sets of $G$. 
 
\begin{definition}
Consider a graph $G(V, E)$,  with ${\bf W} = [W_i]_{i \in V}$ a vector of weights associated with the vertices. Glauber dynamics \cite{dynamic} is a Markov chain over $\mathcal{I}(G)$. Suppose that the chain is at state ${\bf X}(n-1)= [X_i(n-1)]_{i \in V}$. The next transition of Glauber dynamics follows the rules:
\begin{itemize}
\item Select a vertex $i \in V$ uniformly at random.
\item If $\forall j \in \mathcal{N}(i)$, $X_j(n-1) = 0$, then
\begin{displaymath}
	X_i(n) = \left \{ \begin{array}{ll} 
					          1 & \textrm{with probability $\frac{\exp(W_i)}{1+\exp(W_i)}$ } \\
					          0 & \textrm{otherwise.}
					          \end{array} 
					          \right.
\end{displaymath}
\item Otherwise, $X_i(n) = 0$.
\end{itemize}
\end{definition}

The Glauber dynamics is irreducible, aperiodic and time-reversible over $\mathcal{I}(G)$ \cite{dynamic}. It has a product-form stationary distribution, which is:
\begin{equation}
\label{eq:dynamics_sd}
\pi({\bf X}) = \frac{1}{Z}\exp(\sum_{i \in {\bf X}}{W_i}); {\bf X} \in \mathcal{I}(G),
\end{equation}
where $Z$ is a normalizing constant in order to have the sum of the probabilities unit total mass. 

\subsection{Mixing Time}
Glauber dynamics can converge to its stationary distribution starting from any initial distribution. To characterize the convergence speed, we need to quantify the time that it takes for Glauber dynamics to reach close to its stationary distribution.  To establish the result, we need to define some notation first. 

\begin{definition}
(Distance of probability distributions) Given two probability distributions \textrm{\boldmath $\mu$} and  \textrm{\boldmath $\nu$}  over a finite space {\boldmath $\Omega$}, the total variance (TV) distance is defined as:
\begin{equation}
||\textrm{\boldmath $\mu$} - \textrm{\boldmath $\nu$}||_{TV} := \frac{1}{2} \sum_{i \in \textrm{\boldmath $\Omega$}}{|\mu(i) - \nu(i)|},
\end{equation}
and the $\chi^2$ distance \cite{aloha0811}, represented as $\| \frac{\textrm{\boldmath $\mu$}}{\textrm{\boldmath $\nu$}} - 1\|_{2, \textrm{\boldmath $\mu$}}$, is defined as:
\begin{equation}
\Big\|\frac{\textrm{\boldmath $\nu$}}{\textrm{\boldmath $\mu$}} - 1\Big\|^2_{2, \textrm{\boldmath $\mu$}} :=\Big\|\textrm{\boldmath $\nu$} - \textrm{\boldmath $\mu$}\Big\|^2_{2, \frac{1}{\textrm{\boldmath $\mu$}}} = \sum_{i \in \textrm{\boldmath $\Omega$}}{\mu(i)\Big( \frac{\nu(i)}{\mu(i)} - 1 \Big)^2}.
\end{equation}
For any two vectors, {\boldmath $\mu$},  {\boldmath $\nu$} $\in $$\mathbb{R} ^{| \textrm{\boldmath $\Omega$} |}_+$, we define:
\begin{equation}
\| \textrm{\boldmath $\nu$} \|^2_{2, \textrm{\boldmath $\mu$}} := \sum_{i \in \textrm{\boldmath $\Omega$}}{\mu_i \nu_i^2}.
\end{equation}
\end{definition} 

Following the definition, the probability distances satisfy the following condition.

\begin{lemma}
\label{lem:chap12_dine}
\cite{aloha0811} Given two probability distributions \textrm{\boldmath $\mu$} and  \textrm{\boldmath $\nu$}  over a finite space {\boldmath $\Omega$}, the equation below holds:
\begin{equation}
\Big\|\frac{\textrm{\boldmath $\nu$}}{\textrm{\boldmath $\mu$}} - 1\Big\|_{2, \textrm{\boldmath $\mu$}}  \geq 2||\textrm{\boldmath $\nu$} - \textrm{\boldmath $\mu$}||_{TV} 
\end{equation}
\end{lemma}

\begin{IEEEproof}
\begin{eqnarray}
\Big\|\frac{\textrm{\boldmath $\nu$}}{\textrm{\boldmath $\mu$}} - 1\Big\|_{2, \textrm{\boldmath $\mu$}} &=& \sqrt{ \sum_{i \in \textrm{\boldmath $\Omega$}}{\mu(i)\Big( \frac{\nu(i)}{\mu(i)} - 1 \Big)^2}} \nonumber \\
& = & \sqrt{\sum_{i \in \textrm{\boldmath $\Omega$}}{\mu(i)}} \sqrt{ \sum_{i \in \textrm{\boldmath $\Omega$}}{\mu(i)\Big( \frac{\nu(i)}{\mu(i)} - 1 \Big)^2}} \nonumber \\
& \geq & \sum_{i \in \textrm{\boldmath $\Omega$}}{\mu(i) | \frac{\nu(i)}{\mu(i)} - 1 |} \nonumber \\
&& \textrm{  (using Cauchy-Schwarz inequality)} \nonumber \\
& = & \sum_{i \in \textrm{\boldmath $\Omega$}}{|\nu(i) - \mu(i)|} = 2 ||\textrm{\boldmath $\nu$} - \textrm{\boldmath $\mu$}||_{TV} 
\end{eqnarray}
\end{IEEEproof}

\begin{definition}
\label{def:chap12_mixing}
\cite{dynamic} (Mixing time) For a Markov chain with a transition probability matrix {\bf P} and a stationary distribution $\textrm{\boldmath $\pi$}$, define the distance:
\begin{equation}
d(\tau) := \max_{\textrm{\boldmath $\mu$}(0)} { \| \textrm{\boldmath $\mu$}(0) {\bf P}^{\tau} - \textrm{\boldmath $\pi$} \| _{TV} }.
\end{equation}
The mixing time is defined as:
\begin{equation}
\tau_{mix}(\delta) := \min\{ \tau: d(\tau) \leq \delta \}.
\end{equation}
\end{definition}

From the definition, we can see that mixing time is a parameter to measure the convergence rate of a Markov chain to its stationary distribution. Also, following Lemma \ref{lem:chap12_dine}, the mixing time can be measured by calculating the $\chi^2$ distance of $\textrm{\boldmath $\mu$}(\tau)$ and {\boldmath $\pi$} .


\section{DISQUO: A Distributed Algorithm for a Crosspoint Buffered Switch}
\label{sec:disquo}

In this section, we will present DISQUO for a crosspoint buffered switch. The algorithm is totally distributed. Inputs and outputs utilize the states of crosspoint buffers to implicitly exchange information. We will prove the system stability for any admissible Bernoulli traffic, and evaluate the delay performance by running extensive simulations for different traffic patterns.

\begin{figure}[t]
		\centering
    \includegraphics[width=2.6in]{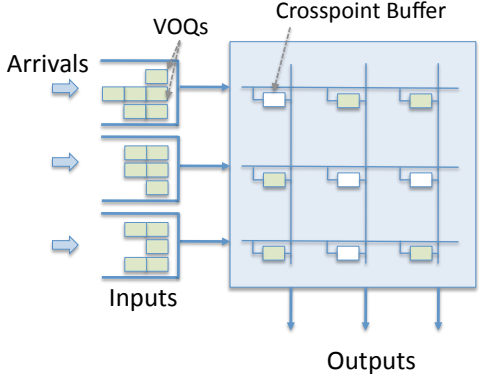}
    \caption{An example of a crosspoint buffered switch. Each input has $N$ \emph{virtual output queues} (VOQs). There is a buffer with a size of $K$ at each crosspoint of the fabric.}
    \label{fig:buffered}
\end{figure}

\subsection{Crosspoint Buffered Switch}
With today's ASIC technology, it is now possible to add a small buffer at each crosspoint inside the crossbar (see Fig. \ref{fig:buffered}). This makes the \emph{crosspoint buffered} or \emph{combined input and crossbar queueing} (CICQ) switch a much more attractive architecture since its scheduler is potentially much simpler. The input and output schedulers can be independent. First, each input picks a crosspoint buffer to send a packet to. Then, each output picks a crosspoint buffer to transmit a packet from. However, existing algorithms \cite{rr_rr, lqf_rr, shen07, shen10} either cannot guarantee $100\%$ throughput or require a centralized scheduler.  

An $N\times N$ crosspoint buffered switch is shown in Fig. \ref{fig:buffered}. We assume fixed size packet (cell) switching. Variable size packets can be segmented into cells before switching and reassembled at the output ports. There are \emph{virtual output queues (VOQs)} at the inputs to prevent head-of-line blocking. Each input maintains $N$ VOQs, one for each output. Let $VOQ_{ij}$ represent the VOQ at input $i$ for output $j$, and $Q_{ij}(n)$ the queue length of $VOQ_{ij}$ at time $n$. Let $(i, j)$ represent the crosspoint between input $i$ and output $j$. 

Each crosspoint has a buffer of size $K$. Most current implementations are constrained by the buffer size. However, it turns out that $K=1$ is sufficient for DISQUO. We will therefore assume that $K=1$ in the following. Our algorithm can be easily extended to the case when $K > 1$. Let $CB_{ij}$ denote the buffer at the crosspoint between input $i$ and output $j$. $B_{ij}(n)$ $\in$ $\{0, 1\}$ denotes the occupancy of $CB_{ij}$ at time $n$.

A schedule can be represented by \textbf{S}(n) = $[\textbf{S}^I(n), \textbf{S}^O(n)]$. $\textbf{S}^I(n) = [S^I_{ij}(n)]$ is the input schedule. Each input port can only transmit at most one cell at each time slot. Thus the input schedule is subject to the following constraints:
\begin{equation}
\label{eq:input_schedule}
\sum_{j}S^I_{ij}(n) \leq 1, \ S^I_{ij}(n) = 0 \textrm{ if }B_{ij}(n)=1.
\end{equation}

$\textbf{S}^O(n) = [S^O_{ij}(n)]$ is the output schedule. It has to satisfy the following constraints:
\begin{equation}
\label{eq:output_schedule}
\sum_{i}S^O_{ij}(n) \leq 1, \ S^O_{ij}(n) = 0 \textrm{ if }B_{ij}(n)=0.
\end{equation}

Let $\lambda_{ij}$ represent the arrival rate of traffic between input $i$ and output $j$. We assume that the arrival process is i.i.d. Bernoulli.
\begin{definition}
An arrival process is said to be \emph{admissible} if it satisfies:
\begin{equation}
\label{eq:admissible}
\sum_j{\lambda_{ij}} < 1 \textrm{, and } \sum_i{\lambda_{ij}} < 1.
\end{equation}
\end{definition}

Let $\textrm{\boldmath $\sigma$}^*$ denote the traffic that the equivalence in Eq. (\ref{eq:admissible}) holds. It is easy to verify that for any admissible traffic {\boldmath $\sigma$}, there exists an $\epsilon > 0$ such that $\textrm{\boldmath $\sigma$} < (1-\epsilon)\textrm{\boldmath $\sigma$}^*$ component-wise. 

Let $||{\bf Q}||$ represent the norm of matrix {\bf Q}: $||{\bf Q}|| = \big( \sum_{i, j}{Q_{ij}^2} \big)^{1/2}$. The stability of a system is defined as:

\begin{definition}
A system of queues is said to be \emph{stable} if:
\begin{equation}
\label{eq:stable}
\lim_{n \to \infty} \sup{E||{\bf Q}(n)||} < \infty.
\end{equation}
\end{definition}

\begin{thm}
\label{thm:equal}
A scheduling algorithm, which can stabilize the system for any admissible traffic in a bufferless crossbar switch, can also stabilize the system for any admissible traffic in a crosspoint buffered switch \cite{shen10}.
\end{thm}

\begin{IEEEproof}
Please refer to Property $1$ of Ref. \cite{shen10}.
\end{IEEEproof}

Following Theorem \ref{thm:equal}, all the scheduling algorithms that have been proposed for an input-queued switch, e.g., the \emph{maximum weight matching} (MWM) \cite{mwm01}, can be applied to a crosspoint buffered switch. As we will show later, the reason that DISQUO can stabilize the system for any admissible traffic is that, after the system converges, the schedule generated at every time slot has a weight that approaches the one with the maximum weight matching schedule. 

\subsection{DISQUO Scheduling Algorithms}
Before presenting the algorithm, we need to introduce some further notation. A DISQUO schedule ${\bf X}(n)$ is a schedule that is generated by the DISQUO algorithm. It is used to determine the input schedules and output schedules. A DISQUO schedule has the following properties: 

\begin{property}
\label{def:disquo}
A {\it DISQUO schedule} ${\bf X}(n)$ can be represented by an $N \times N$ matrix, where $X_{ij}(n) \in \{0, 1\}$, and $\sum_i X_{ij}(n) \leq 1$,
$\sum_j X_{ij}(n) \leq 1$.
\end{property}

With some abuse of notation, we also use ${\bf X}$ to represent a set, and write $(i, j)$ $\in$ ${\bf X}$ if $X_{ij}=1$. Note that a DISQUO schedule {\bf X} has the property that if $X_{ij}=1$, then $\forall i'\ne i $, $X_{i'j}=0$ and $\forall j'\ne j $, $X_{ij'}=0$. We define these crosspoints as its \emph{neighbors} as follows.
\begin{definition}
The neighbors of a crosspoint $(i, j)$ are defined as:
\begin{equation}
\mathcal{N}(i, j) = \{ (i', j)\ or\ (i, j')\ |\ \forall i' \ne i,\  \forall j' \ne j  \}
\end{equation}
\end{definition}

A DISQUO schedule {\bf X} then has the following property:
\begin{property}
If $(i, j)$ $\in$ {\bf X}, $\forall (k, l) \in \mathcal{N}(i, j)$, $(k, l) \notin $ {\bf X}.
\end{property}

Let $\mathcal{X}$ represent the set of all DISQUO schedules. 
\begin{property}
\label{prop:disquo}
At each time slot, when a DISQUO schedule is generated, each input and output port determine their schedules by observing the following rules:
\begin{itemize}
\item For input $i$, when $X_{ij}(n) = 1$, if $Q_{ij}(n) > 0$ and $B_{ij}(n-1)=0$, $S^I_{ij}(n)=1$. Otherwise, $S^I_{ij}(n)=0$.
\item For output $j$, if $X_{ij}(n) = 1$ and $B_{ij}(n) > 0$, $S^O_{ij}(n)=1$.
\end{itemize}
\end{property}

\begin{property}
For an input $i$, if $\forall j$, $X_{ij} = 0$, then it is referred to as a \emph{free input}. A free input port has the freedom to pick any eligible crosspoint to serve, i.e. it can transfer a packet to any empty crosspoint buffer.
\end{property}

\begin{property}
\label{prop:free}
For an output port $j$, if $\forall i$, $X_{ij} = 0$, then it is refered a \emph{free output}. A free output is free to pick any non-empty crosspoint to serve.
\end{property}

Following Prop. \ref{prop:disquo}-\ref{prop:free}, the input schedule ${\bf S}^I(n)$ and output schedule ${\bf S}^O(n)$ can be determined after the DISQUO schedule ${\bf X}(n)$ is generated. Therefore, we will next present the basic DISQUO schedule updating algorithm that generates ${\bf X}(n)$. 

At the beginning, set the initial DISQUO schedule ${\bf X}(0)$ to any schedule that satisfies Property \ref{def:disquo}. For simplicity, we can set ${\bf X}(0) = \emptyset$. At the beginning of a time slot $n$, generate an input/output permutation ${\bf H}(n)$ randomly. Then, the DISQUO schedule ${\bf X}(n)$ is updated following the rules below:

\vspace{0.05in}
{\bf Basic DISQUO Algorithm}

\vspace{0.05in}
\begin{tabular}{p{3.2in}}
\hline
\vspace{0.02in}
$\circ$ $\forall$ $(i, j) \notin {\bf H}(n)$:\\
\hspace{0.2in}
(a) $X_{ij}(n) = X_{ij}(n-1)$.\\
$\circ$ For $(i, j) \in {\bf H}(n)$:\\
\hspace{0.2in} - If $(i, j) \in {\bf X}(n-1)$:\\
\hspace{0.4in}	(b) $X_{ij}(n) = 1$ with probability $p_{ij}$; \\
\hspace{0.4in}	(c) $X_{ij}(n) = 0$ with probability $\overline{p}_{ij} = 1 - p_{ij}$. \\
\hspace{0.2in} - Else, if $(i, j) \notin {\bf X}(n-1)$, and $\forall (k, l) \in \mathcal{N}(i, j)$, \\
\hspace{0.3in} $X_{kl}(n-1) = 0$, then:\\
\hspace{0.4in}	 (d) $X_{ij}(n) = 1$ with probability $p_{ij}$; \\
\hspace{0.4in}	(e) $X_{ij}(n) = 0$ with probability $\overline{p}_{ij} = 1 - p_{ij}$. \\
\hspace{0.2in} - Else, $(i, j) \notin {\bf X}(n-1)$ and $\exists (k, l) \in \mathcal{N}(i, j)$ \\
\hspace{0.3in}  such that $X_{kl}(n-1) = 1$:\\
\hspace{0.4in}	(f) $X_{ij}(n) = 0$. \\
\hline
\end{tabular}

\vspace{0.05in}

$p_{ij}$ is defined as: $p_{ij} = \frac{\exp(W_{ij}(n))}{1+\exp(W_{ij}(n))}$, where $W_{ij}(n)$ is a weight function of the queue size $Q_{ij}(n)$, which is defined as 
\begin{equation}
W_{ij}(n) = f(\tilde{Q}_{ij}(n)).
\end{equation}
$f(\cdot)$ is a concave function which we will define later, $Q_{max}(n) = \max_{i, j}{Q_{ij}(n)}$, and $\tilde{Q}_{ij}(n) = \max \{  f^{-1}(\frac{\epsilon}{2N^2} f(Q_{max}(n)) ), Q_{ij}(n) \} $. Recall that for any admissible traffic {\boldmath $\sigma$}, there exists an $\epsilon > 0$ such that $\textrm{\boldmath $\sigma$} < (1-\epsilon)\textrm{\boldmath $\sigma$}^*$ component-wise. Thus, $\epsilon$ is a small positive number that satisfies the condition $\textrm{\boldmath $\sigma$} < (1-\epsilon)\textrm{\boldmath $\sigma$}^*$. 

Note that in our algorithm, $X_{ij}(n)$ can change only when $(i, j)$ is in ${\bf H}(n)$. Therefore, at every time slot, only $(i, j) \in {\bf H}(n)$ can join or leave the DISQUO schedule. 

Comparing the algorithm with the updating rules of Glauber dynamics, we can see that, DISQUO schedule essentially is a multiple-update version of Glauber dynamics, with a vector of time-varying weights since the weight is a function of the queue length, which changes over time. At every time slot, instead of picking only one VOQ, multiple VOQs are picked by ${\bf H}(n)$ to update their states. Like the Glauber dynamic, ${\bf X}(n)$ also only depends on ${\bf X}(n-1)$. Thus, the DISQUO schedules ${\bf X}(0), {\bf X}(1), \cdots$ form a Markov chain. As we will show later, this Markov chain is irreducible, positive recurrent and time-reversible. 

After DISQUO schedule ${\bf S}(n)$ is generated, inputs and outputs can update their schedules ${\bf S}^I(n)$ and ${\bf S}^O(n)$, and start the packet transmissions. As we will prove later, following this algorithm, the system is stable for any admissible Bernoulli i.i.d. traffic. 

\subsection{Discussions}
As we presented in the previous section, the decision of making a crosspoint $(i, j)$ active or not is based on a probability $p_{ij}$, which depends on not only the local queue sizes, but also a global information $Q_{max}(n)$. However, since $\frac{\epsilon}{2N^2}$ is very small, we can use $W_{ij}(n) = f(Q_{ij}(n))$ directly for real implementation. The introduction of $Q_{max}(n)$ is primarily for technical reasons. Therefore, we have the following conjecture. 

\begin{conj}
DISQUO scheduling algorithm with the weight function defined as $W_{ij}(n) = f(Q_{ij}(n))$ can still stable the system for any admissible traffic.
\end{conj}

To be precise, we still need some message passing between linecards. As suggested in \cite{aloha08, aloha0811}, a rough estimate of $Q_{max}(n)$ is sufficient to guarantee the convergence of the system. Therefore, a low-rate Ethernet connection, which is typical in nowadays router design for backplane control message passing, can be used for linecards to broadcast their local maximum queue sizes so that other linecard can estimate the value of $Q_{max}(n)$. At time slot $kN + i$, only linecard $i$ broadcasts its local maximum queue size. Since at every time slot, there is at most one packet departure/arrival from/to an input, the estimation of ${Q}_{max}(n)$, denoted as $\tilde{Q}_{max}(n)$, satisfies: $Q_{max}(n) - 2N \leq \tilde{Q}_{max}(n)  \leq {Q}_{max}(n) + 2N$. This is sufficient for the system stability. For details, please refer to Ref. \cite{aloha08, aloha0811}.

Recall that we need to generate an input/output permutation ${\bf H}(n)$ \emph{randomly} at every time slot to update the DISQUO schedule. For  the simplicity of practical implementation, we can generate the schedule following a Hamiltonian walk \cite{shah02}, instead of using a totally random algorithm. For a switch of size $N$, there are $N!$ input/output permutations. The Hamiltonian walk schedule ${\bf H}(n)$ visits each of the $N!$ distinct permutations exactly once during every $N!$ slots in a deterministic periodic schedule. This can be simply implemented with a time complexity of $O(1)$ \cite{shah02}. 

In the following, we will show how this conjecture can be implemented in a distributed way, without message passing. We will show the performance of the switch by running extensive simulations. 

\subsection{Distributed Implementation}

As shown in the basic DISQUO scheduling algorithm, ${\bf X}(n)$ is generated based on ${\bf X}(n-1)$. Therefore, each input $i$ needs to keep track of the DISQUO schedule in the previous slot, i.e. for which output $j$ was $X_{ij}(n-1)=1$. Similarly, each output needs to keep track of for which input $i$ was $X_{ij}(n-1)=1$. Since the algorithm is distributed, there is no message passing between inputs and outputs. DISQUO needs to make sure that the inputs and outputs keep a consistent view of the DISQUO schedule. For example, if $X_{ij}(n)=1$, both input $i$ and output $j$ should be aware of this. 

Since the decision for a crosspoint to join or leave the DISQUO schedule needs the queue length information, inputs are responsible for making the decisions. However, there are two problems that have to be solved: 1) before input $i$ decides to change $X_{ij}$ from $ 0$ to $1$, it needs to check the states of all the neighbors of $(i, j)$, namely, the status of output $j$, which is not directly accessible at input $i$; 2) after input $i$ changes the value of $X_{ij}$, this information has to be passed over to output $j$. To solve the problems, the distributed DISQUO algorithm is designed to achieve \emph{implicit} message passings by utilizing the crosspoint buffers. 

The input and output scheduling algorithms work as follows. 

\vspace{0.05in}

\begin{tabular}[ht]{p{3.2in}}
{\bf Input Scheduling Algorithm (ISA)}\\
\hline \vspace{0.02in}
At each input port $i$, assume $(i, j)$ $\in$ $\bf {H}(n)$.\\
1) $\forall j' \neq j$: (a) $X_{ij'}(n) = X_{ij'}(n-1)$.\\
2) $\circ$ If $X_{ij}(n-1)=1$:\\
\hspace{0.22in}	(b) $X_{ij}(n) = 1$ with probability $p_{ij}$; \\
\hspace{0.22in}	(c) $X_{ij}(n) = 0$ with probability $\overline{p}_{ij} = 1 - p_{ij}$. \\
\hspace{0.1in} $\circ$ Else, if $X_{ij}(n-1)=0$ and there exists a $j'$ such\\
\hspace{0.12in} that $X_{ij'}(n-1)=1$, $X_{ij'}(n)=1$ according to case \\
\hspace{0.12in} (a) above. Consequently:\\
\hspace{0.22in}  (d) $X_{ij}(n)=0$.\\
\hspace{0.1in}  $\circ$ Else, if there is no $j'$ such that $X_{ij'}(n-1)=1$,\\
\hspace{0.12in} then input $i$ was a free input:\\
\hspace{0.22in} - If $CB_{ij}$ is empty, output $j$ was free: \\
\hspace{0.42in}	(e) $X_{ij}(n) = 1$ with probability $p_{ij}$; \\
\hspace{0.42in}	(f) $X_{ij}(n) = 0$ with probability $\overline{p}_{ij} = 1 - p_{ij}$. \\
\hspace{0.22in} - Else, \\
\hspace{0.42in}	(g) $X_{ij}(n) = 0$.\\
3) If $X_{ij}(n)=1$, $Q_{ij}(n) > 0$ and $B_{ij}(n) = 0$, $S^I_{ij}(n)=$ \\ 
\hspace{0.08in} $1$. Input $i$ sends a packet to $CB_{ij}$. Otherwise, if input \\
\hspace{0.08in}  $i$  is free, it generates ${\bf H}(n+1)$. Suppose that $(i, j')$ \\
\hspace{0.08in}  $\in {\bf H}(n+1)$. If $CB_{ij'}$ is empty, input $i$ serves it. \\
\hspace{0.08in} Otherwise, it sends a packet to any empty crosspoint  \\
\hspace{0.08in} buffer except $CB_{ij}$. \\
\hline
\end{tabular}

\vspace{0.05in}

\begin{tabular}[t]{p{3.2in}}
{\bf Output Scheduling Algorithm (OSA)}\\
\hline\vspace{0.02in}
For each output port $j$, assume $(i, j)$ is selected by ${\bf H}(n)$.\\
1) $\forall i' \neq i$: (a) $X_{i'j}(n) = X_{i'j}(n-1)$.\\
2) $\circ$ If $X_{ij}(n-1)=1$:\\
\hspace{0.2in} (b) If at time $n$, $CB_{ij}$ receives a packet from input \\
\hspace{0.3in} $i$, $X_{ij}(n) = 1$.\\
\hspace{0.2in} (c) Otherwise, $X_{ij}(n) = 0$. \\
\hspace{0.1in} $\circ$ Else, if $X_{ij}(n-1)=0$ and there exists a $i'$ such\\
\hspace{0.1in} that  $X_{i'j}(n-1) =1$, $X_{i'j}(n) =1$. So:\\
\hspace{0.2in} (d) $X_{ij}(n) = 0$. \\
\hspace{0.1in} $\circ$ Else, there is no $i'$ such that $X_{i'j}(n-1)=1$, output\\
\hspace{0.1in} $j$ was free:\\
\hspace{0.2in} - If input $i$ sends a packet to $CB_{ij}$ at the beginning\\
\hspace{0.3in} of time $n$: \\
\hspace{0.5in}	(e) $X_{ij}(n)=1$.\\
\hspace{0.2in} - Else, \\
\hspace{0.5in}	(f) $X_{ij}(n)=0$.\\
3) If $X_{ij}(n)=1$, $S^O_{ij}(n)=1$. Output $j$ transmits a\\
\hspace{0.08in} packet from $CB_{ij}$. Otherwise, output $j$ is free, it\\
\hspace{0.08in} generates  ${\bf H}(n+1)$. Suppose that $(i', j) \in {\bf H}(n+1)$. \\
\hspace{0.08in}  If $CB_{i'j}$ is non-empty, output $j$ serves it. Otherwise, \\
\hspace{0.08in}  output $j$ picks any non-empty crosspoint to serve. \\
\hline
\end{tabular}
\vspace{0.05in}

For the input scheduling algorithm, before input $i$ decides to change $X_{ij}$ from $0$ to $1$, it has to check the status of output $j$. But this information is not known at input $i$. If input $i$ was free at time $n-1$, it has to generate ${\bf H}(n)$. Suppose that $(i, j) \in {\bf H}(n)$. Input $i$ then transmits a packet to $CB_{ij}$ if $CB_{ij}$ was empty at time $n-1$. As we show in the output scheduling algorithm below, if output $j$ was free, it transmits a packet from $CB_{ij}$ at time $n-1$. Therefore, for case (e) and (f), input $i$ can infer whether output $j$ was free or not by observing the crosspoint buffer $CB_{ij}$. If the $CB_{ij}$ is empty, output $j$ was free at time $n-1$. Otherwise, output $j$ was busy.

For the output schedulers, an output $j$ has to observe the crosspoint buffer $CB_{ij}$ to learn input $i$'s decision following the output scheduling algorithm (OSA) above. For example, in case (b) and (c), it can learn input $i$'s decision by observing whether a packet is sent to $CB_{ij}$ or not at time $n$. Thus, when $(i, j) \in {\bf H}(n)$, a packet transmission from input $i$ to $CB_{ij}$ can implicitly pass its decision information to output $j$. If $CB_{ij}$ is not empty at the beginning of time $n$, input $i$ is not able to pass its decision information to output $j$. Recall that if output $j$ was free, it can pick any non-empty crosspoint buffer to serve. Therefore, if output $j$ was free at time $n-1$, it can calculate ${\bf H}(n)$ in advance, and transmit the packet from $CB_{ij}$, if $(i, j) \in {\bf H}(n)$ and $CB_{ij}$ is not empty. By doing that, when $(i, j) \in {\bf H}(n)$, $CB_{ij}$ will always be empty at the beginning of time $n$ if output j was free, so that input $i$ can pass its decision information to output $j$ by sending a packet to the buffer. With this rule, when $(i, j) \in {\bf H}(n)$, input $i$ can also directly infer that output $j$ was busy if $CB_{ij}$ is not empty at the beginning of time $n$, since otherwise, output $j$ would have already transmitted that packet from $CB_{ij}$ at time $n-1$.

\subsection{Example}

\begin{figure}[t]
\centering
    \includegraphics[width=3.4in]{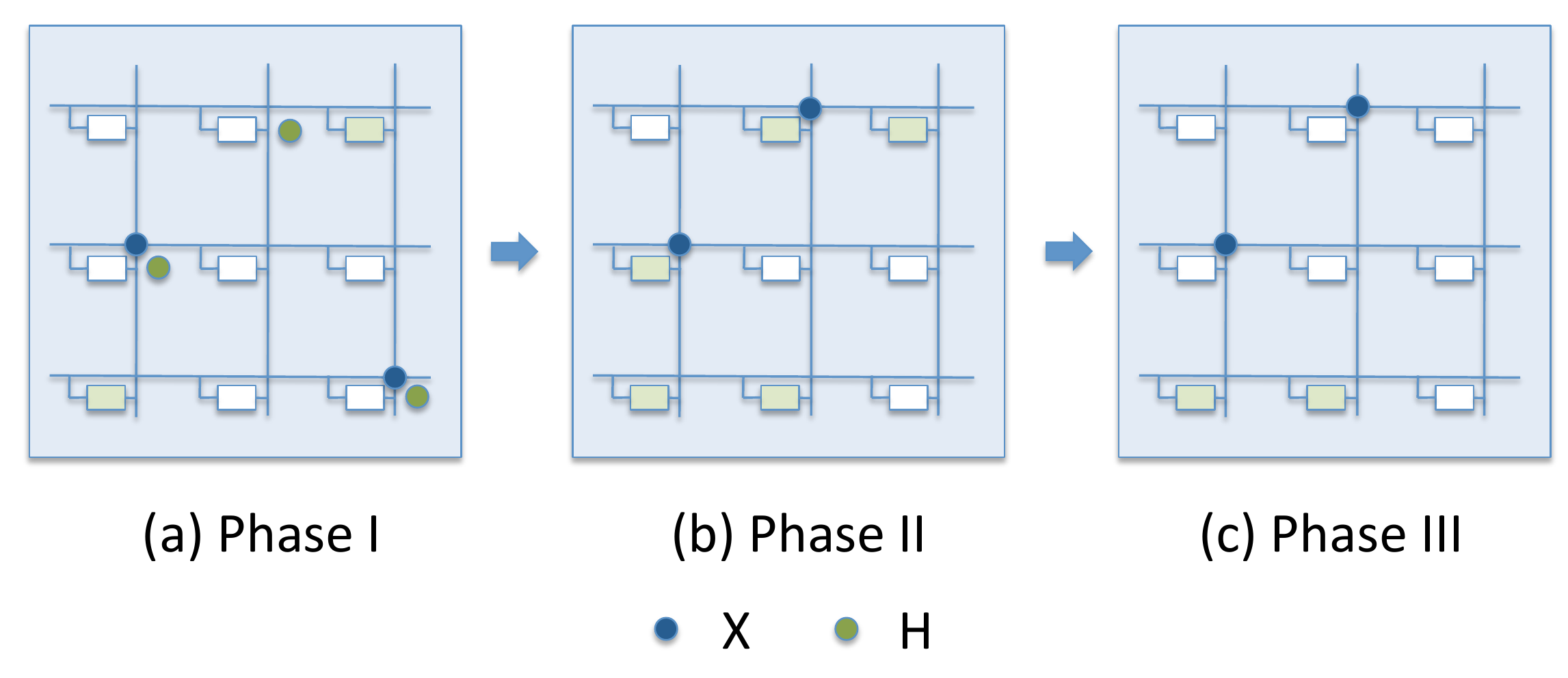}
    \caption{An example of DISQUO schedule updating. One time slot is divided into three phases. ${\bf H}(n)$ is generated in Phase I to update ${\bf X}(n-1)$. In Phase II, inputs make their decisions and transmits packets to the crosspoint buffers. In Phase III, outputs updates their schedules and transmits packets from the crosspoint buffers. Only $(i, j) \in {\bf H}(n)$ can join or leave the DISQUO schedule. }
    \label{fig:exam}
\end{figure}

To help understand DISQUO, we give an illustrative example here. We assume that a schedule over one time slot can be divided into three phases: a) Phase I: every input and output calculate the Hamiltonian walk schedule ${\bf H}(n)$; b) Phase II: inputs update the DISQUO schedule based on ${\bf H}(n)$, and decide the value of ${\bf S}^I(n)$, after which packets can be sent from inputs to the crosspoint buffers; c) Phase III: outputs update the DISQUO schedule and decide the value of ${\bf S}^O(n)$ so that they can transmit packets from the crosspoint buffers. 
		
	As we can see from Fig. \ref{fig:exam}(a), the DISQUO schedule at time $n-1$ is ${\bf S}(n-1) = \{ (2, 1), (3, 3)\}$. In Phase I, $\textbf{H}(n)$ = $\{(1, 2), (2, 1), (3, 3)\}$ is generated at each input and output. In the following, we use the example to describe how a crosspoint joins or leaves the DISQUO schedule, and how the input/output scheduler ${\bf S}^I(n)$ and ${\bf S}^O(n)$ are decided after ${\bf X}(n)$ is generated.
	\begin{itemize}
	\item How a crosspoint joins the DISQUO schedule: $(1, 2)$ is in ${\bf H}(n)$ and $X_{12}(n-1) = 0$. Also, input $1$ knows that $\forall j$, $X_{1j}(n-1) = 0$ so that input 1 was a free input in the previous slot. If output $2$ was also a free output, input $1$ can decide whether to let $(1, 2)$ join the DISQUO schedule or not, following case (e) or (f) of ISA. Input $1$ cannot know the status of output $2$ directly. However, it can learn output $2$'s status by observing $CB_{12}$. Since $CB_{12}$ is empty, input $i$ learns that output $2$ was free. It can then decide whether to make $(1, 2)$ active based on $p_{12}$. If its decision is to set $X_{12}(n)$ to $1$, it should send a packet to $CB_{12}$. Otherwise, it remains a free input. In the example, the decision of input $1$ is to set $X_{12}(n)$ to $1$. Thus, $S^I_{12}(n)=1$, and it sends a packet to $CB_{12}$, as shown in Fig. \ref{fig:exam}(b). Note that this transmission implicitly passes its decision information to output $2$. 
	
	Output $2$ was a free output, and it observes that in Phase II, input $1$ sends a packet to $CB_{12}$. Following case (e) of OSA, output $2$ learns input $1$'s decision of setting $X_{12}(n)$ to $1$.  It then updates $X_{12}(n)$ to $1$, and thus $S^O_{12}(n) =1$. Output $2$ transmits the packet from $CB_{12}$, which is shown in Fig. \ref{fig:exam}(c). 
		
	\item How a crosspoint leaves the DISQUO schedule: $(3, 3)$ is in ${\bf H}(n)$ and $X_{33}(n-1) = 1$. Following case (b) and (c), input $3$ has to decide whether to keep $(3, 3)$ in the DISQUO schedule or not, based on a probability $p_{33}$ which is a function of the queue size $Q_{33}$. In the example, it decides to set $X_{33}(n)$ to $0$. Input $3$ becomes a free input. It calculates ${\bf H}(n+1)$, which we assume is $\{(1, 3), (2, 1), (3, 2)\}$. Since $(3, 2) \in {\bf H}(n+1)$ and $CB_{32}$ is empty, it sets $S^I_{32}(n) = 1$, and sends a packet to $CB_{32}$ (Fig. \ref{fig:exam}(b)). Note that by not sending a packet to $CB_{33}$, input $3$ implicitly passes its decision of setting $X_{33}(n)=0$ to output $3$. 
	
	Output 3 observes that, in Phase II, input $3$ did not send any packet to $CB_{33}$. Following case (b), it learns input $3$'s decision and updates $X_{33}(n)$ to $0$. Output $3$ becomes a free output. Following the OSA, a free output has to generate ${\bf H}(n+1)$ at time $n$. Since $(1, 3) \in {\bf H}(n+1)$ and $CB_{13}$ is not empty, output $3$ sets $S^O_{13}(n)=1$ and transmits the packet from $CB_{13}$, as shown in Fig. \ref{fig:exam}(c). 
	
	\end{itemize}
	
From the example, we can see that, after ${\bf X}(n)$ is generated, which is $\{ (1,2), (2, 1)\}$, a packet is transmitted from input $1$ to output $2$, and one from input $2$ to output $1$. Besides that, input $3$ and output $3$, which are free, also transmit a packet. The transmissions by free inputs and outputs can be considered as an \emph{augmentation} of ${\bf X}(n)$. In the following, we will show that the weight, only defined on ${\bf X}(n)$, is close enough to the maximum one to guarantee the throughput. The augmenting by free input/output, though it does not contribute to the stability of the switch, can improve the delay performance of the system. 	
	
\subsection{Stationary Distribution}	
As mentioned before, $\{ {\bf X}(n)\}$ forms a Markov chain. In this section, we will derive the stationary distribution of this Markov chain, and show that after the system converges, the weight of the DISQUO schedule is always very close to the MWM schedule in Lemma \ref{lem:wc}. We can then prove the system stability in Theorem \ref{thm:stability}.
\begin{lemma}
If ${\bf X}(n-1)$ $\in$ $\mathcal{X}$, then ${\bf X}(n)$ $\in$ $\mathcal{X}$.
\end{lemma}
\begin{IEEEproof}
If {\bf X} is a DISQUO schedule it satisfies Property \ref{def:disquo}. For an input $i$, it is impossible that there exists $j \neq j'$ such that $X_{ij}(n) = X_{ij'}(n) = 1$, since before input $i$ decides to change $X_{ij}$ from $0$ to $1$, it always has to make sure that there does not exist a $j'$ such that $X_{ij'}(n) = 1$. 

For an output $j$, it is also impossible that there exists $i \neq i'$ such that $X_{ij}(n) = X_{i'j}(n) = 1$. This is because input $i$ can change $X_{ij}$ from $0$ to $1$ only when output $j$ was free. So, $X_{ij}(n) = X_{i'j}(n) = 1$ only when input $i$ and input $i'$ decide to change the values from $0$ to $1$ at the same time slot, which requires both $(i, j) \in {\bf H}(n)$ and $(i', j) \in {\bf H}(n)$. But ${\bf H}(n)$ is an input/output permutation such that only one $(\cdot, j)$ is in ${\bf H}(n)$. Therefore, if ${\bf X}(n-1)$ satisfies Property 1, ${\bf X}(n)$ also satisfies Property \ref{def:disquo}.
\end{IEEEproof}

As mentioned before, ${\bf X}(n-1), {\bf X}(n), \cdots$ is a Markov chain since ${\bf X}(n)$ only depends on ${\bf X}(n-1)$. A transition from a state ${\bf X}$ to ${\bf X}'$ can occur only when the Hamiltonian walk schedule ${\bf H}(n)$ satisfies the condition:
\begin{displaymath}
({\bf X} \cap \overline{{\bf X}'}) \cup (\overline{{\bf X}} \cap {\bf X}') \in {\bf H}(n).
\end{displaymath}
This is because VOQs in ${\bf X} \cap \overline{{\bf X}'}$ leave the DISQUO schedule and $\overline{{\bf X}} \cap {\bf X}'$ join the DISQUO schedule. According to the DISQUO algorithm, only crosspoints in ${\bf H}(n)$ can join or leave the DISQUO schedule. Therefore, both ${\bf X} \cap \overline{{\bf X}'}$  and $\overline{{\bf X}} \cap {\bf X}'$ should be in ${\bf H}(n)$. The following lemma gives the transition probabilities. 

\begin{lemma}
\label{lem:tran_prob}
If ${\bf X}$ can transit to ${\bf X}'$, the transition probability can be written as:
\begin{eqnarray}
\label{eq:tran_prob}
p_n(\textbf{X}, \textbf{X}') &&= \sum_{ \textbf{H}: {\bf X \bigtriangleup X' \in H}}{
a(\textbf{H})
\prod_{(i, j) \in {\bf X \cap \overline{X}'}} {\overline{p}_{ij}}
\prod_{(k, l) \in {\bf \overline{X} \cap X'} } {p_{kl}} }{}\nonumber\\
&&{} \cdot
\prod_{(u,v) \in {\bf{X} \cap X' \cap H}} {p_{uv}}
\prod_{(x,y) \in {\bf H \cap \overline{X \cup X'}} \cap \overline{\mathcal{N}({\bf X \cup X'})}} {\overline{p}_{xy}}
,{}\nonumber\\
\end{eqnarray}
where $a(\textbf{H})$ is the probability that \textbf{H} is selected (which is $\frac{1}{N!}$), and ${\bf X} \bigtriangleup {\bf X}'$ = $({\bf X} \cap \overline{{\bf X}'}) \cup (\overline{{\bf X}} \cap {\bf X}')$.
\end{lemma}

\begin{IEEEproof}
Please refer to Appendix \ref{ap:tran_prob}.
\end{IEEEproof}

\begin{lemma}
\label{lm:recurrent}
The Markov chain $\{{\bf X}(n)\}$ is irreducible and positive recurrent.
\end{lemma}

\begin{IEEEproof}
Please refer to Appendix \ref{ap:recurrent}.
\end{IEEEproof}

Since the Markov chain is positive recurrent, it has a unique stationary distribution. Let us associate each VOQ with a non-negative weight $W_{ij}(n) = f(Q_{ij}(n))$ at time $n$. The Markov chain has the following stationary distribution. 

\begin{lemma}
The Markov chain of the system has the following product-form stationary distribution:
\begin{equation}
\label{eq:sd}
\pi_n({\bf X}) = \frac{1}{\mathcal{Z}} \prod_{(i,j)\in {\bf X}}{\frac{p_{ij}}{\overline{p}_{ij}}}
= \frac{1}{\mathcal{Z}} \prod_{(i,j)\in {\bf X}}{e^{W_{ij}(n)}} ,
\end{equation}
where
\begin{equation}
\label{eq:sdz}
\mathcal{Z} = \sum_{{\bf X} \in \mathcal{X}}{\prod_{(i,j)\in {\bf X}}{\frac{p_{ij}}{\overline{p}_{ij}}}}
= \sum_{{\bf X} \in \mathcal{X}}{\prod_{(i,j)\in {\bf X}}{e^{W_{ij}(n)}}}.
\end{equation}
\end{lemma}

\begin{IEEEproof}
If a state {\bf X} can make a transition to ${\bf X}'$, we can check that the distribution in Eq. (\ref{eq:sd}) satisfies the detailed balance equation:
\begin{equation}
\label{eq:balance}
\pi_n ({\bf X}) p_n({\bf X},{\bf X}') = \pi_n ({\bf X}') p_n({\bf X}',{\bf X}),
\end{equation}
hence the Markov chain is reversible and Eq. (\ref{eq:sd}) is the stationary distribution (see \cite{kelly}, Theorem 1.2).
\end{IEEEproof}

\subsection{System Convergence}
For Glauber dynamics, the weights are fixed over time. Therefore, the convergence rate of the system can be described by the distance between {\boldmath $\mu$(n)} and {\boldmath $\pi$}. However, following the algorithm presented in the previous section, the weights are changing over time such that the Glauber dynamics for each time slot $n$ is different from those in other time slots, which means $\textrm{\boldmath $\pi$}_n$ also varies over time. To characterize the convergence rate of this system, we can define the distance 
\begin{equation}
d_n = \| \textrm{\boldmath $\mu$}_n - \textrm{\boldmath $\pi$}_n \|_{TV}.
\end{equation}
We aim to ensure that for any arbitrarily small $\delta > 0$, there exists a time $T_{mix}(\delta)$ that for any $n > T_{mix}(\delta)$, we have $d_n < \delta$ so that $\textrm{\boldmath $\mu$}_n$ and $\textrm{\boldmath $\pi$}_n$ are close enough. As compared to the definition of mixing time in Definition \ref{def:chap12_mixing}, $T_{mix}(\delta)$ shows the convergence rate of the system. Therefore, we will refer to $T_{mix}(\delta)$ as the mixing time of this inhomogeneous Markov chain. 

In the following Lemma, we will prove that if the weight function $f(x)$, so that $W_{ij}(n) = f(Q_{ij}(n))$ ,  is carefully selected, the system can always converge to the distribution $\textrm{\boldmath $\pi$}_n$ following DISQUO scheduling algorithm. 

\begin{lemma}
\label{lem:convergence}
If $f(x) = \frac{\log(1+x)}{g(x)}$, then there exists a $n^*$ that for any $\delta > 0$, $\| \textrm{\boldmath $\mu$}_n - \textrm{\boldmath $\pi$}_n \|_{TV} \leq \delta$ holds for all $n \geq n^*$, where $g(x)$ is a function that satisfies the following conditions:
\begin{itemize}
	\item $g(x) \geq 1$, for all $x \geq 0$.
	\item $g'(x) \geq 0$, for all $x \geq 0$.
	\item $\lim_{x \to \infty } g(f^{-1}(x)) = \infty$.
\end{itemize} \end{lemma}

\begin{IEEEproof}
Please refer to Appendix \ref{sec:appb} for the detailed proof.
\end{IEEEproof}
One example of $g(x)$ is $g(x) = \log(e+\log(1+x))$. Note that according to Lemma \ref{lem:convergence}, if the weight function is well designed, the system will always converge to the product-form distribution as expressed in Eq. (\ref{eq:sd}).

\subsection{System Stability}
As shown above, the Markov chain $\{{\bf X}(n)\}$ has a finite number of states and we already derived its stationary distribution. In the following, we will utilize MWM algorithm to prove the system stability.  For an input-queued switch, the \textbf{MWM} algorithm selects a feasible schedule ${\bf S}(n)$ with the maximum weight:
\begin{equation}
\label{eq:mwm}
\textbf{S}^{*}(n) = \arg\max_{{\bf S} \in \mathcal{S}} \sum_{(i,j) \in {\bf S}}{W_{ij}(n)}.
\end{equation}

The algorithm can provide $100\%$ throughput for any admissible traffic in a bufferless crossbar switch. According to Theorem \ref{thm:equal}, MWM can be extended to a buffered crossbar switch. Following the DISQUO algorithm, if $X_{ij}(n) = 1$, and $Q_{ij}(n)>0$ or $B_{ij}(n)>0$, one packet can be transmitted from input $i$ to output $j$. Therefore, we can define the weight of a DISQUO schedule as:
\begin{equation}
\label{eq:ps_weight}
W({\bf X}) = \sum_i{\sum_j{X_{ij}(n) W_{ij}(n)}}.
\end{equation}

For \textbf{MWM}, the result below has been established in \cite{ton05}.
\begin{lemma}
\label{lm:stable_cond}
For a scheduling algorithm, if given any $\epsilon$ and $\delta$ such that $0 \le \epsilon$, $\delta < 1$, there exists a $B > 0$ such that the scheduling algorithm satisfies the condition that in any time slot $n$, with a probability greater than $1 - \delta$, the scheduling algorithm can choose a feasible schedule {\bf S} which satisfies the following condition:
\begin{equation}
\label{eq:mwm2}
\sum_{(i,j) \in {\bf S}(n)}{W_{ij}(n)} \geq (1-\epsilon)\sum_{(k,l) \in {\bf S}^{*}(n)}{W_{kl}(n)},
\end{equation}
whenever $||\textbf{Q}(n)|| \geq B$, where $\textbf{Q}(n) = [Q_{ij}(n)]_{i, j}$ and $||\textbf{Q}(n)|| = \big(\sum_{i,j}{Q_{ij}^2(n)}\big)^{1/2}$. Then the scheduling algorithm can stabilize the system.
\end{lemma}

\begin{thm}
\label{thm:stability}
DISQUO can stabilize the system if the input traffic is admissible.
\end{thm}

\begin{IEEEproof}
Define the set:
\begin{displaymath}
\mathcal{K} = \{ {\bf X} \in \mathcal{X}: W({\bf X}) \leq  (1 - \epsilon)W^*({\bf X}) \}.
\end{displaymath}
According to Lemma \ref{lem:wc} in Appendix \ref{ap:stability}, for any $\delta > 0$, we have $\pi(\mathcal{K}) < \delta$, if the maximum weight satisfies the condition:
\begin{equation}
\label{eq:mw_cond}
W^*({\bf X}) > \frac{N^2 \log 2}{\epsilon \delta} > \frac{\log |\mathcal{X}|}{\epsilon \delta}.
\end{equation}

So, for any $\epsilon, \delta > 0$, there exists a $B > N^3 \frac{\log2}{\epsilon \delta}$ such that whenever $||\textbf{Q}(n)|| > B$,
\begin{displaymath}
\sum_{i,j}{Q_{ij}^2(n)} > B^2 > N^6 \big(\frac{\log2}{\epsilon \delta} \big)^2.
\end{displaymath}
Then, $\max{Q_{ij}^2(n)} > N^4 \big(\frac{\log 2}{\epsilon \delta} \big)^2$. Thus, Eq. (\ref{eq:mw_cond}) holds and $\pi(\mathcal{K}) < \delta$. Hence the scheduling algorithm can stabilize the system according to Lemma \ref{lm:stable_cond}.
\end{IEEEproof}


\section{Simulations}
\label{sec:simulation}
\begin{figure}[tb]
		\centering
    \includegraphics[width=2.8in]{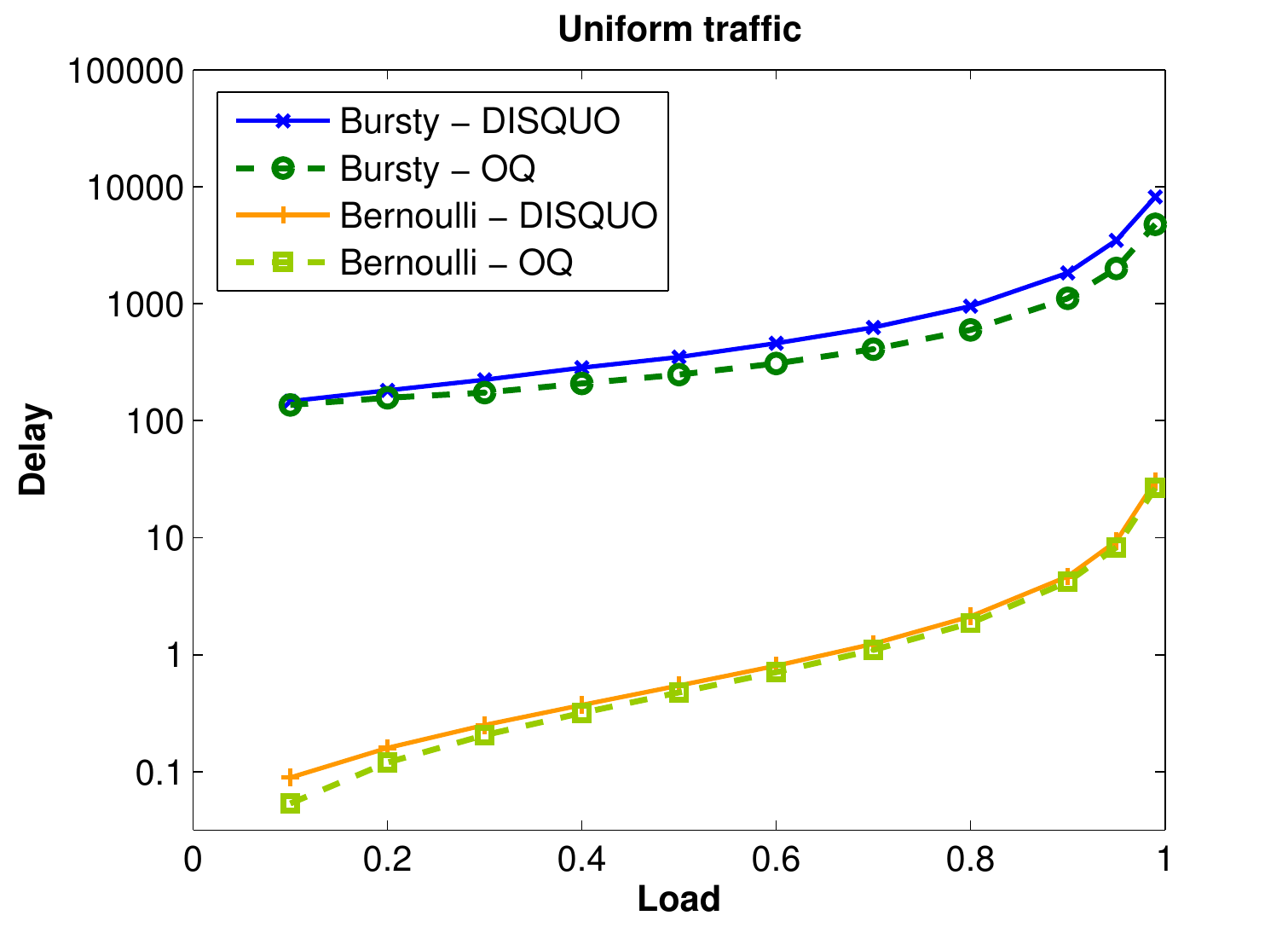}
    \caption{Switch size N=32, uniform traffic for both Bernoulli i.i.d. and bursty arrivals}
    \label{fig:dis_uniform}
\end{figure}

In this section, we ran simulations for different scenarios to evaluate the performance of DISQUO. We also study the delay performance of the scheduling algorithm under different traffic patterns, including uniform and non-uniform traffic with Bernoulli and bursty arrivals. Note that DISQUO reduces to a heuristic scheduling algorithm for all arrival processes that are not i.i.d. Bernoulli. For bursty traffic, the burst length is distributed over $[1, 1000]$, following the truncated Pareto distribution:
\begin{equation}
\label{eq:bursty}
P(l) = \frac{c}{l^\alpha},\ l=\textrm{1, 2, ... , 1000},
\end{equation}
where $l$ is the burst length, $\alpha$ is the Pareto distribution parameter and $c$ is the normalization constant. In the simulations, $\alpha=1.7$, for which the average burst length is about $11.6$. All inputs are equally loaded and we measure the packet delay. Simulations are run for long enough to ensure that the confidence intervals are small enough to make valid comparisons. 

\subsection{Uniform Traffic}
\begin{figure}[tb]
		\centering
    \includegraphics[width=2.8in]{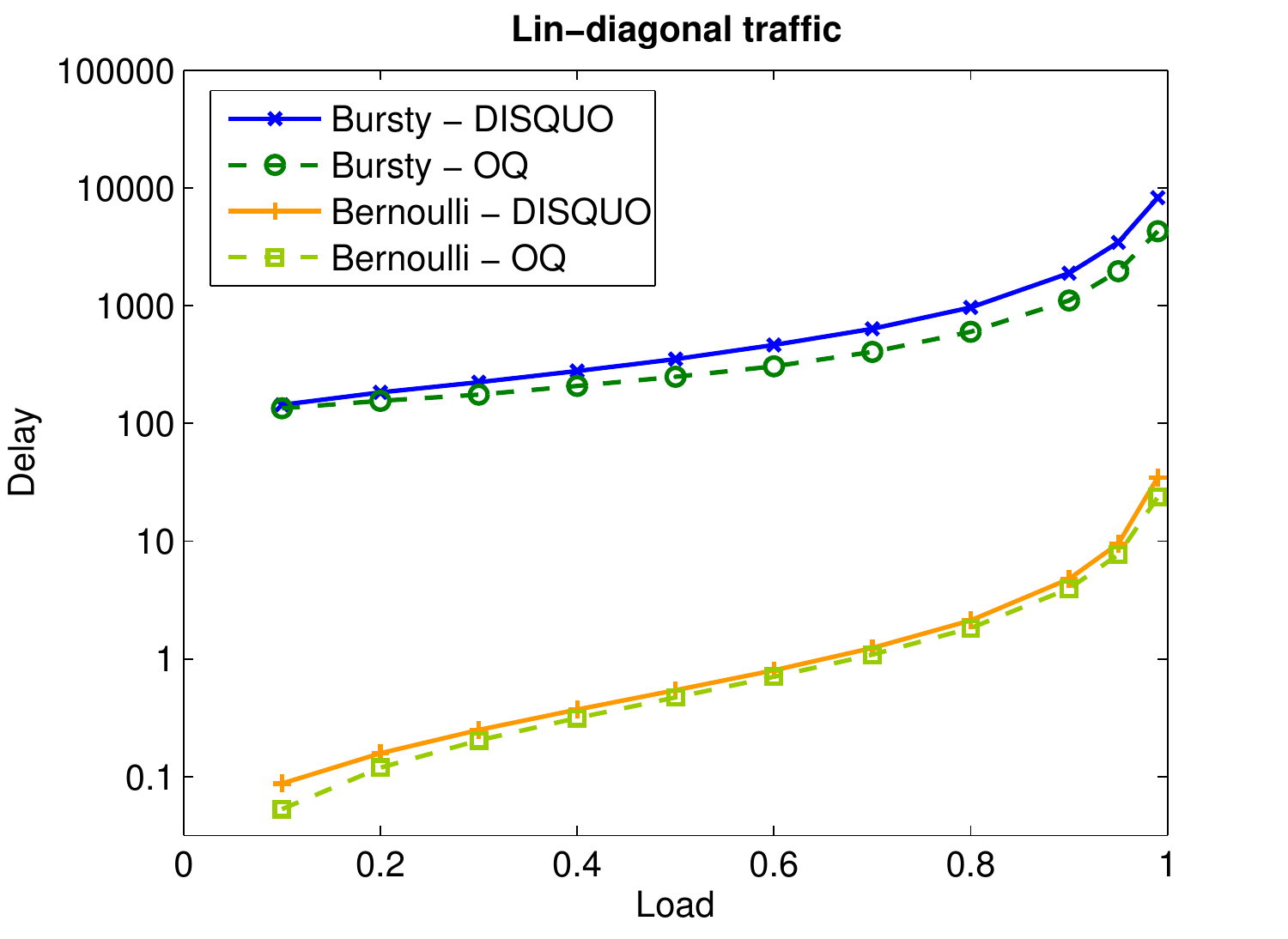}
    \caption{Switch size N=32, lin-diagonal traffic for both Bernoulli i.i.d. and bursty arrivals}
    \label{fig:dis_lin}
\end{figure}

For uniform traffic, a new cell is destined with equal probability to all output ports. Let $\sigma$ represent the traffic load, and the arrival rate between input $i$ and output $j$ is $\sigma_{ij} = \frac{\sigma}{N}$. The delay performance of DISQUO under uniform Bernoulli and bursty traffic is shown in Fig. \ref{fig:dis_uniform}. We can see that the packet delay of DISQUO is very close to the output-queued switch (OQ). It has been shown that under uniform traffic, even an algorithm as simple as RR-RR can have a delay performance close to an output-queued switch \cite{rr_rr}. However, the RR-RR algorithm cannot achieve $100\%$ throughput when the traffic is non-uniform. Therefore, we will study the performance of DISQUO under non-uniform traffic next.

\subsection{Non-uniform Traffic}

\begin{figure}[tb]
		\centering
    \includegraphics[width=2.8in]{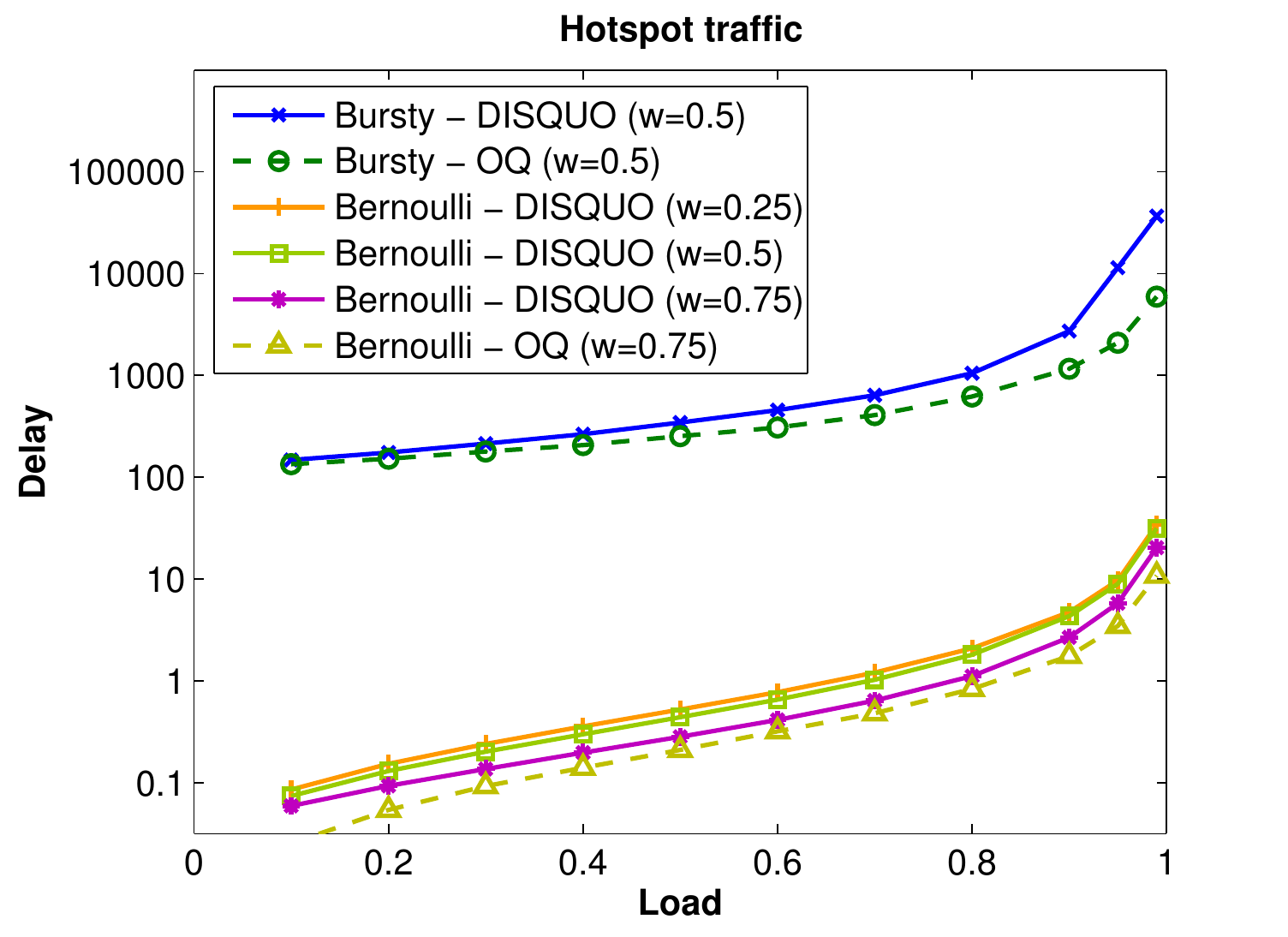}
    \caption{Switch size N=32, hot-spot traffic for both Bernoulli i.i.d. and bursty arrivals}
    \label{fig:dis_hot}
\end{figure}

We ran the simulations for the following non-uniform traffic patterns:
\begin{itemize}
	\item Lin-diagonal: Arrival rates at the same input differ linearly, i.e, $\sigma_{i(i+j\pmod{N})} - \sigma_{i(i+j+1\pmod{N})} = 2 \sigma /N(N+1)$.
	\item Hot-spot: For input port $i$, $\sigma_{ii} = \omega \sigma$ and $\sigma_{ij} = (1-\omega) \sigma/(N-1)$, for $i \neq j$. We can get different traffic patterns by varying the hot-spot factor $\omega$.
\end{itemize}

The delay performance for lin-diagonal and hot-spot traffic are shown in Fig. \ref{fig:dis_lin} and Fig. \ref{fig:dis_hot}, respectively. We can see that under Bernoulli traffic, the delay performance of DISQUO is still very close to the output-queued switch. Packets have low delay even when the load is as high as $0.99$. Note that the RR-RR algorithm can have a throughput of approximately only $85\%$ \cite{rr_rr} under hotspot traffic. Note that DISQUO is stable for the bursty traffic scenarios that we simulated.

\begin{figure}[t]
		\centering
    \includegraphics[width=2.8in]{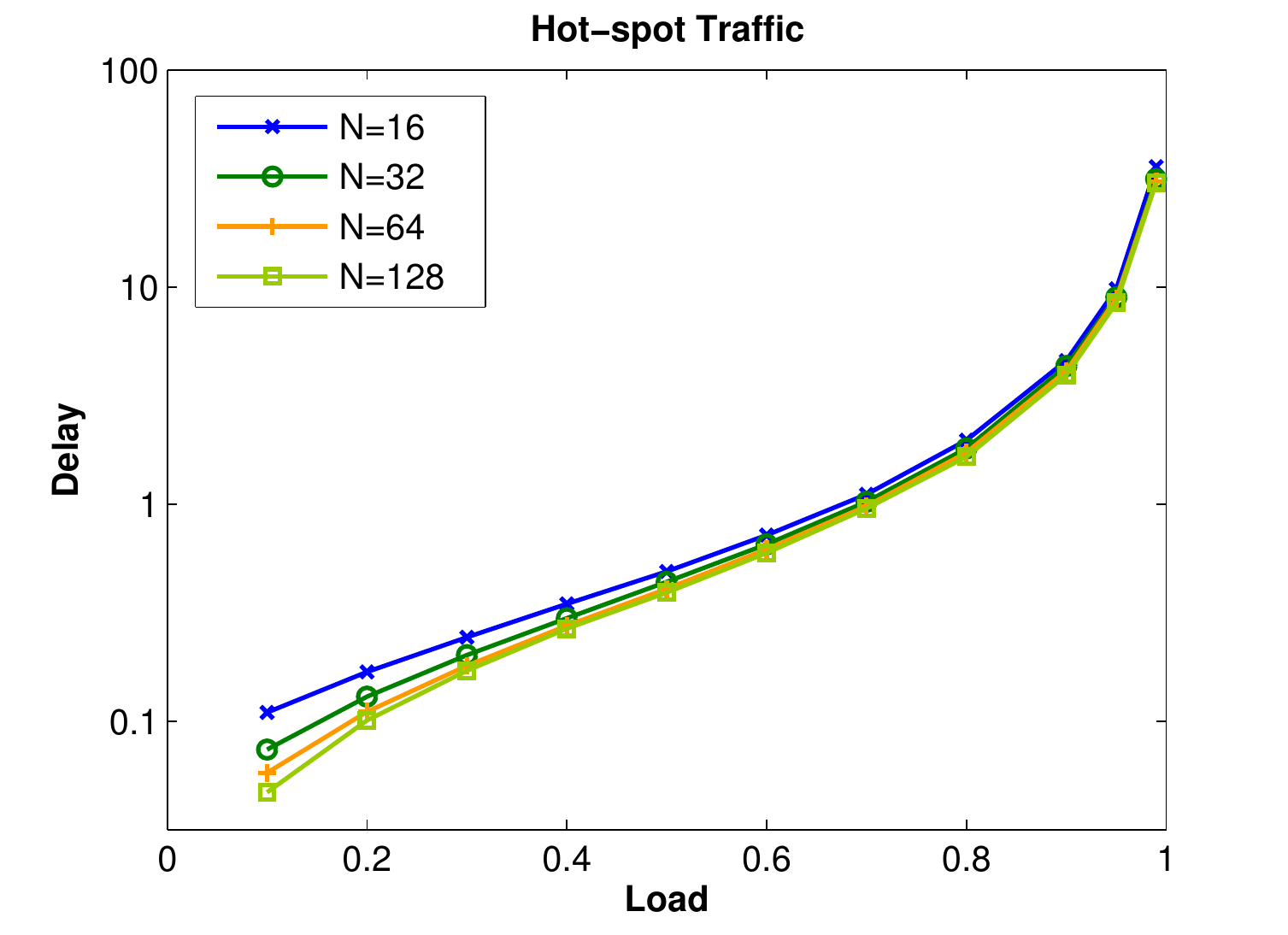}
    \caption{Results of switches with different sizes, with hot-spot Bernoulli i.i.d. traffic, where $\omega=0.5$}
    \label{fig:dis_size}
\end{figure}

\subsection{Impact of Switch Size}

We also study the impact of switch size on the delay performance. Generally, for input-queued switches, the average delay increases linearly with the switch size \cite{islip}. For output-queued switches, delay is independent of the size. Fig. \ref{fig:dis_size} shows the delay performance of DISQUO with different switch sizes under Bernoulli hot-spot traffic, for which $\omega$ is $0.5$. We can see that the delay is almost the same for different switch sizes. As the size increases, the delays even decrease slightly. This is partly because as the switch size increases, the number of crosspoint buffers increases as well, and the crosspoint buffers play a key role in reducing the average delay. 

\subsection{Impact of Buffer Size}
\begin{figure}[t]
		\centering
    \includegraphics[width=2.8in]{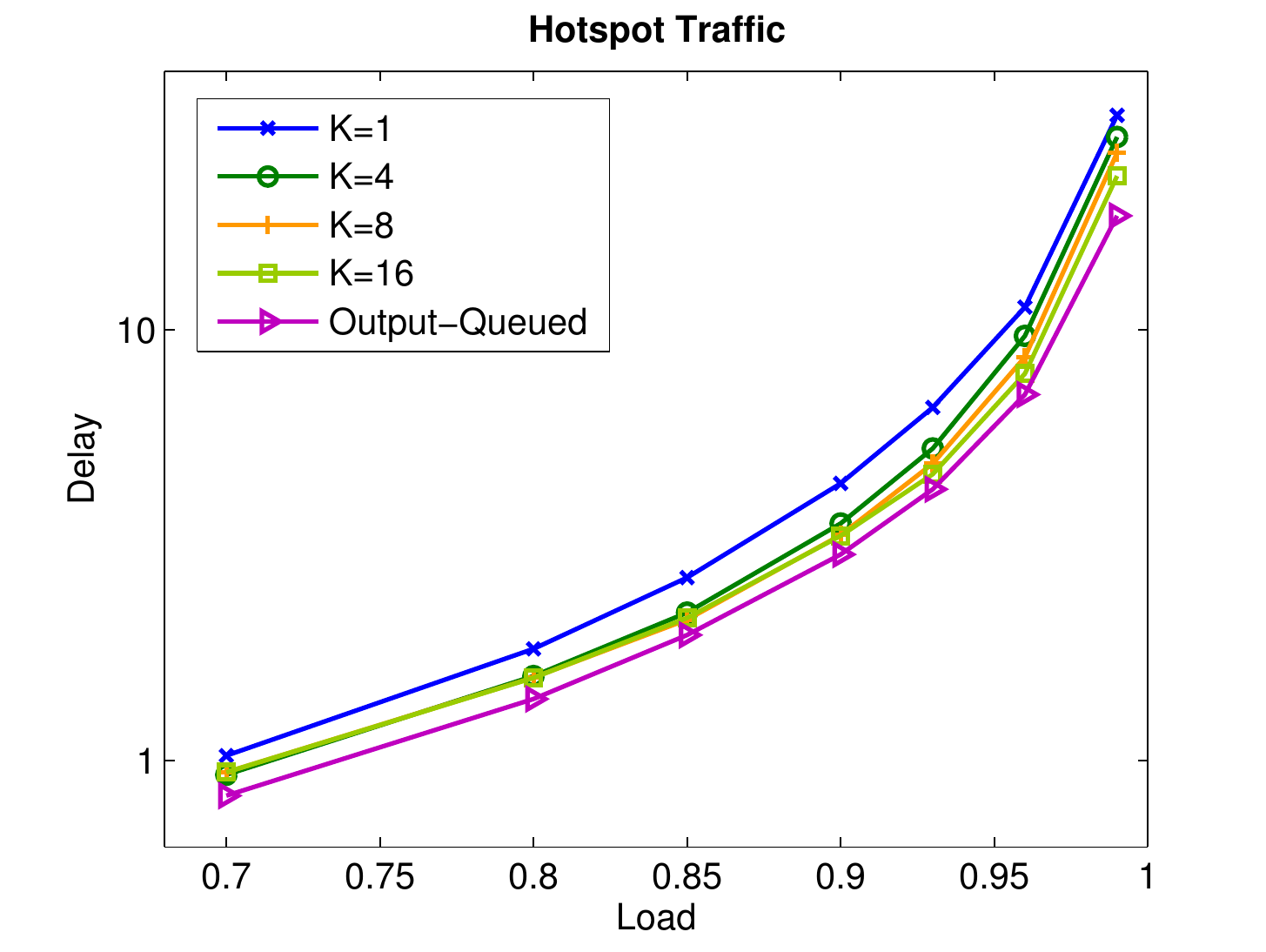}
    \caption{Impact of buffer size, hot-spot traffic, $\omega=0.5$, N=32}
    \label{fig:buffer}
\end{figure}

If the buffer at each crosspoint increases to infinity, the buffered crossbar switch is then equivalent to an output-queued switch. So if we increase the buffer size, the average delay will decrease and converge to the delay of an output-queued switch. As we already showed in previous simulation results, the delay performance of DISQUO with a buffer size of $1$ is already very close to that of an output-queued switch. Therefore, by increasing the buffer size, we can only get a very marginal improvement in delay performance. DISQUO can be easily modified for values of $K>1$. Due to space considerations, we will not define DISQUO with $K>1$ here. Fig. \ref{fig:buffer} shows the delay performance of DISQUO with different buffer sizes, under hot-spot traffic. We can see that the improvement is small. Therefore, we only need to implement a one-cell buffer at each crosspoint and still provide good delay performance. This is crucial since current technology limits the size of crosspoint buffers to a small number.

\section{Conclusion}
\label{sec:conclusion}
In this paper, we first proposed a distributed scheduling algorithm (DISQUO) for crosspoint buffered switches with a crosspoint buffer size of as small as one and no speedup. The computational complexity of DISQUO is only $O(1)$ per port, and we proved that it can achieve $100\%$ throughput for any admissible Bernoulli i.i.d. traffic. We evaluated the performance of DISQUO by running extensive simulations. The results show that DISQUO can provide very good delay performance, as compared to an output-queued switch. With DISQUO, the average queuing delay for a packet is independent of the switch size, which makes it very suitable for large-scale switching system design.

\bibliographystyle{ieeetr}
\bibliography{journal}

\begin{thebibliography}{10}

\bibitem{hem}
R.~Hemenway, R.~Grzybowski, C.~Minkenberg, and R.~Luijten, ``{Optical-
  packet-switched Interconnect for Supercomputer Applications},'' {\em Journal
  of Optical Networks}, vol.~3, no.~12, pp.~900--913, 2004.

\bibitem{mink}
C.~Minkenberg, F.~Abel, R.~Krishnamurthy, M.~Gusat, P.~Dill, I.~Iliadis,
  R.~Luijten, B.~R. Hemenway, R.~Grzybowski, and E.~Schiattarella, ``{Designing
  a Crossbar Scheduler for HPC Applications},'' {\em IEEE Micro}, vol.~26,
  pp.~58--71, May-June 2006.

\bibitem{fares}
M.~Al-Fares, A.~Loukissas, and A.~Vahdat, ``{A Scalable, Commodity, Data Center
  Network Architecture},'' in {\em Proceedings of ACM SIGCOMM}, August 2008.

\bibitem{helios}
N.~Farrington, G.~Porter, S.~Radhakrishnan, H.~H. Bazzaz, V.~Subramanya,
  Y.~Fainman, G.~Papen, and A.~Vahdat, ``{Helios: A Hybrid Electrical/Optical
  Switch Architecture for Modular Data Centers},'' in {\em Proceedings of ACM
  SIGCOMM}, Aug-Sep 2010.

\bibitem{mwm01}
L.~Tassiullas and A.~Ephremides, ``{Stability Properties of Constrained Queuing
  Systems and Scheduling Policies for Maximum Throughput in Multihop Radio
  Networks},'' {\em IEEE Transactions on Automatic Control}, vol.~37,
  pp.~1936--1949, December 1992.

\bibitem{mwm02}
N.~McKeown, A.~Mekkittikul, V.~Anantharam, and J.~Walrand, ``{Achieving $100\%$
  Throughput in an Input-Queued Switch},'' {\em IEEE Transactions on
  Communications}, vol.~47, pp.~1260--1267, August 1999.

\bibitem{islip}
N.~Mckeown, ``{The iSLIP Scheduling Algorithm for Input-Queued Switches},''
  {\em IEEE/ACM Transactions on Networking}, vol.~7, pp.~188--201, April 1999.

\bibitem{drrm}
Y.~Li, S.~Panwar, and H.~J. Chao, ``{On the Performance of a Dual Round-Robin
  Switch},'' in {\em Proc. of IEEE INFOCOM}, April 2001.

\bibitem{pim}
T.~E. Anderson, S.~S. Owichi, J.~B. Saxe, and C.~P. Thacher, ``{High Speed
  Switch Scheduling for Local Area Networks},'' {\em ACM Transactions on
  Computer Systems}, vol.~11, pp.~319--352, November 1993.

\bibitem{maximal}
J.~G. Dai and B.~Prabhakar, ``{The Throughput of Data Switches with and without
  Speedup},'' in {\em Proceedings of IEEE INFOCOM}, (Tel Aviv, Israel), 2000.

\bibitem{cioq}
S.~Chuang, A.~Goel, N.~McKeown, and B.~Prabhakar, ``{Matching Output Queueing
  with a Combined Input Output Queued Switch},'' in {\em Proc. of IEEE
  INFOCOM}, March 1999.

\bibitem{rr_rr}
R.~Rojas-Cessa, E.~Oki, and H.~J. Chao, ``{On the Combined Input-Crosspoint
  Buffered Packet Switch with Round-Robin Arbitration},'' {\em IEEE
  Transactions on Communications}, vol.~53, pp.~1945--1951, November 2005.

\bibitem{lqf_rr}
T.~Javidi, R.~Magill, and T.~Hrabik, ``{A High Throughput Scheduling Algorithm
  for a Buffered Crossbar Switch Fabric},'' in {\em Proceedings of IEEE ICC},
  (Helsinki, Finland), June 2001.

\bibitem{sbf}
L.~Mhamdi and M.~Hamdi, ``{MCBF: a High-Performance Scheduling Algorithm for
  Buffered Crossbar Switches},'' {\em IEEE Communications Letters}, vol.~7,
  pp.~451--453, September 2003.

\bibitem{chuang05}
S.-T. Chuang, S.~Iyer, and N.~McKeown, ``{Practial Algorithms for Performance
  Guarantees in Buffered Crossbars},'' in {\em Proceedings of IEEE INFOCOM},
  (Miami, Florida), March 2005.

\bibitem{turner06}
J.~Turner, ``{Strong Performance Guarantees for Asynchronous Crossbar
  Schedulers},'' in {\em Proceedings of IEEE INFOCOM}, (Spain), April 2006.

\bibitem{shah05}
P.~Giaccone, E.~Leonardi, and D.~Shah, ``{On th emaximal throughput of networks
  with finite buffers and its application to buffered crossbars},'' in {\em
  Proceedings of IEEE Infocom}, (Miami, FL), March 2005.

\bibitem{auction}
M.~Bayati, B.~Prabhakar, D.~Shah, and M.~Sharma, ``{Iterative Scheduling
  Algorithms},'' in {\em Proc. of IEEE INFOCOM}, May 2007.

\bibitem{aloha08}
S.~Rajagopalan, D.~Shah, and J.~Shin, ``{Network adiabatic theorem: an
  efficient randomized protocol for contention resolution},'' in {\em
  Proceedings of ACM SIGMETRICS}, June 2009.

\bibitem{jiang08}
L.~Jiang and J.~Walrand, ``{A Distributed CSMA Algorithm for Throughput and
  Utility Maximization in Wireless Networks},'' {\em IEEE/ACM Transactions on
  Networking}, vol.~18, pp.~960--972, June 2010.

\bibitem{ni0908}
J.~Ni and R.~Srikant, ``{Q-CSMA: Queue-Length Based CSMA/CA Algorithms for
  Achieving Maximum Throughput and Low Delay in Wireless Networks},'' {\em
  Proceedings of IEEE Infocom}, Apr. 2010.

\bibitem{liu10}
J.~Liu, Y.~Yi, A.~Proutiére, M.~Chiang, and H.~V. Poor, ``{Towards
  utility-optimal random access without message passing},'' {\em Wireless
  Communications and Mobile Computing}, vol.~10, pp.~115--128, Jan. 2010.

\bibitem{ni1008}
L.~Jiang, M.~Leconte, J.~Ni, R.~Srikant, and J.~Walrand, ``{Fast Mixing of
  Parallel Glauber Dynamics and Low-Delay CSMA Scheduling},'' {\em submitted},
  August 2010.

\bibitem{ghaderi}
J.~Ghaderi and R.~Srikant, ``{On the Design of Efficient CSMA Algorithms for
  Wireless Networks},'' {\em submitted}, 2010.

\bibitem{dynamic}
D.~A. Levin, Y.~Peres, and E.~L. Wilmer, {\em {Markov Chains and Mixing
  Times}}.
\newblock American Mathematical Society, 2009.

\bibitem{aloha0811}
S.~Rajagopalan and D.~Shah, ``{Aloha that works},'' {\em submitted}, Nov. 2008.

\bibitem{shen07}
Y.~Shen, S.~S. Panwar, and H.~J. Chao, ``{Providing $100\%$ Throughput in a
  Buffered Crossbar Switch},'' in {\em Proceedings of IEEE HPSR}, (Brooklyn,
  New York), May-June 2007.

\bibitem{shen10}
Y.~Shen, S.~S. Panwar, and H.~J. Chao, ``{SQUID: A Practical 100
  Scheduler for Crosspoint Buffered Switches},'' {\em IEEE/ACM Transactions on
  Networking}, vol.~18, pp.~1119--1131, August 2010.

\bibitem{shah02}
P.~Giaccone, B.~Prabhakar, and D.~Shah, ``{Toward Simple, High Performance
  Schedulers for High-Aggregate Bandwidth Switches},'' in {\em Proceedings of
  IEEE INFOCOM}, (New York), 2002.

\bibitem{kelly}
F.~Kelly, {\em {Reversibility and Stochastic Networks}}.
\newblock Wiley, 1979.

\bibitem{ton05}
A.~Eryilmaz, R.~Srikant, and J.~R. Perkins, ``{Stable Scheduling Policies for
  Fading Wireless Channels},'' {\em IEEE/ACM Transactions on Networking},
  vol.~13, pp.~411--424, April 2005.

\bibitem{cbound01}
M.~Jerrum and A.~Sinclair, ``{Approximating the permanent},'' {\em SIAM Journal
  of Computing}, vol.~18, pp.~1149--1178, 1989.

\bibitem{cbound02}
P.~Bremaud, {\em {Markov chains, Gibbs fields, Monte Carlo simulation and
  queues}}.
\newblock Springer-Verlag, 2001.

\end{thebibliography}

\appendix
\subsection{Proof of Lemma \ref{lem:tran_prob}}
\label{ap:tran_prob}
\begin{IEEEproof}
The transition occurs only when the VOQs in \textbf{H} satisfy the conditions below:
\begin{enumerate}
\item For any $(i, j)$ $\in$ ${\bf X} \cap \overline{{\bf X}'}$: the VOQ is selected by \textbf{H} and decides to change its scheduling decision from $1$ to $0$, which happens with probability $\overline{p}_{ij}$.

\item For any $(k, l)$ $\in$ $\overline{{\bf X}} \cap {\bf X}'$: the VOQ is selected by \textbf{H} and decides to change its scheduling decision from $0$ to $1$, which happens with probability $p_{kl}$.

\item For any $(u, v)$ $\in$ ${\bf X} \cap {\bf X}' \cap \textbf{H}$: the VOQ was in the DISQUO schedule of the previous time slot, and even though selected by \textbf{H} it decides to keep its state, which occurs with probability $p_{uv}$.

\item For any $(x, y)$ $\in$ $\textbf{H} \cap \overline{{\bf X} \cup {\bf X}'} \cap \overline{\mathcal{N}({\bf X})}$: neither the VOQ nor any of its neighbors was in the DISQUO schedule of the previous time slot, and though selected by \textbf{H} it decides to keep its schedule, which occurs with probability $\overline{p}_{xy}$. Since \textbf{H} is a DISQUO schedule and $\overline{{\bf X}} \cap {\bf X}' \in \textbf{H}$, $\textbf{H} \cap \mathcal{N}(\overline{{\bf X}} \cap {\bf X}') = \emptyset$. Thus
$\textbf{H} \cap \overline{{\bf X} \cup {\bf X}'} \cap \overline{\mathcal{N}({\bf X})}$ =
$\textbf{H} \cap \overline{{\bf X} \cup {\bf X}'} \cap \overline{\mathcal{N}({\bf X} \cup {\bf X}')}$.
We replace
$\textbf{H} \cap \overline{{\bf X} \cup {\bf X}'} \cap \overline{\mathcal{N}({\bf X})}$
by
$\textbf{H} \cap \overline{{\bf X} \cup {\bf X}'} \cap \overline{\mathcal{N}({\bf X} \cup {\bf X}')}$
in Eq. (\ref{eq:tran_prob}) for the proof of the stationary distribution in the following.
\end{enumerate}

Since $\textbf{H}$ is a permutation of the inputs and outputs, for any two VOQs in {\bf H}, they are not neighbors of each other. Therefore, they can make the scheduling decisions independently. We can then multiply the probabilities of all the four categories above, which leads to the transition probability given by Eq. (\ref{eq:tran_prob}). 
\end{IEEEproof}

\subsection{Proof of Lemma \ref{lm:recurrent}}
\label{ap:recurrent}
\begin{IEEEproof}
Suppose that {\bf X} is a DISQUO schedule, and it has $k$ non-zero elements: $(i_1, j_1)$, $(i_2, j_2)$ $\cdots$ $(i_k, j_k)$ $\in$ {\bf X}. Let ${\bf X}_l$ represent a DISQUO schedule which has $l$ non-zero elements: $(i_1, j_1)$, $(i_2, j_2)$ $\cdots$ $(i_l, j_l)$ $\in$ ${\bf X}_l$ $\subseteq$ {\bf X}, $0\le l \le k$. We can see that ${\bf X}_0 = \textbf{0}$ and ${\bf X}_k = {\bf X}$. Since {\bf X} is a DISQUO schedule, ${\bf X}_l$ is also a DISQUO schedule and ${\bf X}_{l-1} \cup {\bf X}_l = {\bf X}_l \in \mathcal{X}$. Therefore, the system can make a transition from ${\bf X}_{l-1}$ to ${\bf X}_l$ with positive probability when $(i_l, j_l) \in {\bf H}(n)$, as we already proved in Lemma \ref{lem:tran_prob}. Hence, state ${\bf X}_0$ can reach any state {\bf X} $\in$ $\mathcal{X}$ with positive probability in a finite number of steps and vice versa. Thus, the Markov chain is irreducible and positive recurrent. 
\end{IEEEproof}

\subsection{Lemmas for System Stability}
\label{ap:stability}

\begin{lemma}
\label{lm:f_fun}
Suppose that $T(\cdot)$ is a function defined on a set $\mathcal{X}$. For any probability distribution $\mu$ on $\mathcal{X}$, define the function:
\begin{equation}
F(\mu, T({\bf X})) = E_{\mu}[T({\bf X})] + \textrm{$\mathcal{H}$}(\mu),
\end{equation}
where $\mathcal{H}(\mu)$ is the entropy function: $- \sum_{{\bf X} \in \mathcal{X}}{\mu({\bf X}) \log \mu({\bf X})}$. Then $F(\cdot)$ is uniquely maximized by the distribution:
\begin{equation}
\label{eq:max_mu}
\mu ^*({\bf X}) = \frac{1}{Z} \exp (T({\bf X})),
\end{equation}
where $Z = \sum_{{\bf X} \in \mathcal{X}}{\exp (T({\bf X}))}$.
\end{lemma}

\begin{IEEEproof}
For any probability distribution $\mu$, we have:
\begin{eqnarray}
& & F(\mu, T({\bf X}))   {}\nonumber  \\
& = & E_{\mu}[T({\bf X})] + \textrm{$\mathcal{H}$}(\mu) {}\nonumber  \\
			 & = & \sum_{{\bf X} \in \mathcal{X}} { \mu({\bf X}) T({\bf X}) } - \sum_{{\bf X} \in \mathcal{X}}{\mu({\bf X}) \log \mu({\bf X})} {}\nonumber \\
			 & = & \sum_{{\bf X} \in \mathcal{X}} { \mu({\bf X}) (\log \mu ^* ({\bf X}) + \log Z) } - \sum_{{\bf X} \in \mathcal{X}}{\mu({\bf X}) \log \mu({\bf X})} {}\nonumber \\
			 & = & \sum_{{\bf X} \in \mathcal{X}} { \mu({\bf X}) \log Z} + \sum_{{\bf X} \in \mathcal{X}} {\mu({\bf X}) \log \frac{\mu ^* ({\bf X})}{\mu({\bf X})} } {}\nonumber \\
			 & \leq & \log Z \sum_{{\bf X} \in \mathcal{X}} { \mu({\bf X})} + \log \Big( \sum_{{\bf X} \in \mathcal{X}} { \mu({\bf X})\frac{\mu ^* ({\bf X})}{\mu({\bf X})} }\Big)  {}\nonumber \\
			 & = & \log Z,
\end{eqnarray}
with equality holding only when $\mu = \mu^*$. \textbf{QED}
\end{IEEEproof}

Note that when $T({\bf X}) = 0$, the uniform distribution maximizes $F(\mu, 0)$, and we have:
\begin{equation}
\label{eq:entropy}
F(\mu, 0) = \textrm{$\mathcal{H}$}(\mu) \leq \log Z = \log |\mathcal{X}|
\end{equation}
where $|\mathcal{X}|$ is the size of $\mathcal{X}$.

\begin{lemma}
\label{lem:wc}
Let $W(\cdot)$ be the weight function and $W^*({\bf X})$ the maximum weight. Define the set:
\begin{equation}
\label{eq:kk}
\mathcal{K} = \{ {\bf X} \in \mathcal{X}: W({\bf X}) \leq  (1 - \epsilon)W^*({\bf X}) \}.
\end{equation}
Then, we have:
\begin{equation}
\label{eq:kkk}
\pi(\mathcal{K}) \leq \frac{\log |\mathcal{X}|}{\epsilon W^*({\bf X})}
\end{equation}
\end{lemma}

\begin{IEEEproof}
As shown in Eq. (\ref{eq:sd}), for a schedule ${\bf X} \in \mathcal{X}$, its stationary distribution is: $\pi({\bf X}) = \frac{1}{\mathcal{Z}} \prod_{(i, j) \in {\bf X}} {e^{(w_{ij}(n))}} = \frac{1}{\mathcal{Z}} e^{W({\bf X})}$. According to Lemma \ref{lm:f_fun}, $\pi$ maximizes $F(\mu, W({\bf X}))$.

Let ${\bf X}^*$ be the schedule that maximizes the weight, and $\pi'$ be the distribution that assigns all probability on ${\bf X}^*$ such that:
\begin{displaymath}
\pi '({\bf X}) = \left\{
\begin{array}{ll}
1 & \textrm{if {\bf X} = ${\bf X}^*$} \\
0 & \textrm{otherwise}
\end{array} \right.
\end{displaymath}
Then we have:
\begin{eqnarray}
\label{eq:ww}
F(\pi', W({\bf X})) & = & E_{\pi'}[W({\bf X})] + \textrm{$\mathcal{H}$}(\pi') {}\nonumber \\
											 & = & W^*({\bf X}) + \textrm{$\mathcal{H}$}(\pi') {}\nonumber \\
											 & \leq & F(\pi, W({\bf X})) =  E_{\pi}[W({\bf X})] + \textrm{$\mathcal{H}$}(\pi) {}\nonumber \\
											 & \leq & W^*({\bf X})(1- \pi(\mathcal{K}) )  {}\nonumber \\
											 & &  +  W^*({\bf X})(1-\epsilon)\pi(\mathcal{K}) + \textrm{$\mathcal{H}$}(\pi) {}\nonumber \\
											 & = & W^*({\bf X})(1 - \epsilon \pi(\mathcal{K})) + \textrm{$\mathcal{H}$}(\pi)
\end{eqnarray}
The last step in Eq. (\ref{eq:ww}) uses Eq. (\ref{eq:kk}). So,
\begin{eqnarray}
W^*({\bf X}) + \textrm{$\mathcal{H}$}(\pi') & \leq & W^*({\bf X})(1 - \epsilon \pi(\mathcal{K})) + \textrm{$\mathcal{H}$}(\pi) {}\nonumber \\
\epsilon \pi(\mathcal{K})W^*({\bf X}) & \leq & \textrm{$\mathcal{H}$}(\pi)  - \textrm{$\mathcal{H}$}(\pi') \leq \textrm{$\mathcal{H}$}(\pi) \leq \log |\mathcal{X}| {}\nonumber \\
\pi(\mathcal{K}) & \leq & \frac{\log |\mathcal{X}|}{\epsilon W^*({\bf X})}
\end{eqnarray}
\end{IEEEproof}

\subsection{Proof of System Convergence}
\label{sec:appb}

Before presenting the proof, we need to introduce some preliminaries. We will first define a \emph{matrix norm}, which will be useful in determining the mixing time of a finite-state Markov chain.
\begin{definition}
(Matrix norm) Consider a $|\textrm{\boldmath $\Omega$}| \times |\textrm{\boldmath $\Omega$}|$ non-negative matrix  {\boldmath $A$} $\in \mathbb{R}_+^{|\textrm{\boldmath $\Omega$}| \times |\textrm{\boldmath $\Omega$}|}$ and a given vector $\textrm{\boldmath $\mu$} \in \mathbb{R}_+^{|\textrm{\boldmath $\Omega$}|}$. Then, the matrix norm of  {\boldmath $A$} with respect to  {\boldmath $\mu$} is defined as:
\begin{equation}
\| {\bf A} \|_{\textrm{\boldmath $\mu$}} = \sup_{\textrm{\boldmath $\nu$}: E_{\textrm{\boldmath $\mu$}}[\textrm{\boldmath $\nu$}] = 0}{\frac{\| {\bf A} \textrm{\boldmath $\nu$} \| _{2, \textrm{\boldmath $\mu$}}}{ \| \textrm{\boldmath $\nu$} \| _{2, \textrm{\boldmath $\mu$}} }},
\end{equation}
where $\textrm{\boldmath $\nu$} \in \mathbb{R}_+^{|\textrm{\boldmath $\Omega$}|} $ and $E_{\textrm{\boldmath $\mu$}}[\textrm{\boldmath $\nu$}]  = \sum_i{\mu_i \nu_i}$.
\end{definition}

It is easy to check that the matrix norm has the following properties \cite{aloha0811}:
\begin{property}
For a matrix ${\bf A} \in \mathbb{R}_+^{|\textrm{\boldmath $\Omega$}| \times |\textrm{\boldmath $\Omega$}|}$, $\textrm{\boldmath $\pi$} \in \mathbb{R}_+^{| \textrm{\boldmath $\Omega$} |}$ and $a \in \mathbb{R}$:
\begin{equation}
\| a {\bf A} \|_{\textrm{\boldmath $\pi$}} = |a| \| {\bf A} \|_{\textrm{\boldmath $\pi$}}.
\end{equation}
\end{property}

\begin{property}
{\bf A} and {\bf B} are the transition matrices of two reversible Markov chains. They have the same stationary distribution which is $\textrm{\boldmath $\pi$}$. We then have:
\begin{equation}
\|  {\bf A} {\bf B}  \|_{\textrm{\boldmath $\pi$}} \leq  \| {\bf A} \|_{\textrm{\boldmath $\pi$}}  \| {\bf B} \|_{\textrm{\boldmath $\pi$}}.
\end{equation}
\end{property}

\begin{property}
Let {\bf P} be the transition matrix of a reversible Markov chain, which has the stationary distribution $\textrm{\boldmath $\pi$}$. We then have:
\begin{equation}
\|  {\bf P} \|_{\textrm{\boldmath $\pi$}} \leq  e_{\max},
\end{equation}
where $e_{\max} = \max \{ |e|: |e| \neq 1,  \textrm{e is an eigenvalue of {\bf P}} \}$ and $0 < e_{max} < 1$.
\end{property}

With the definition and these properties, it follows that for any distribution {\boldmath $\mu$} on {\boldmath $\Omega$}, , we have  \cite{aloha0811}:
\begin{equation}
\Big\| \frac{\textrm{\boldmath $\mu$} {\bf P} }{\textrm{\boldmath $\pi$} } -1 \Big\| _{2, \textrm{\boldmath $\pi$} } \leq \|{\bf P}^*\|_{\textrm{\boldmath $\pi$}} \Big\| \frac{\textrm{\boldmath $\mu$}}{\textrm{\boldmath $\pi$} } -1 \Big\| _{2, \textrm{\boldmath $\pi$} }.
\end{equation}
Then, if the Markov chain is time-reversible, we have:
\begin{equation}
\Big\| \frac{\textrm{\boldmath $\mu$}(\tau)}{\textrm{\boldmath $\pi$} } -1 \Big\| _{2, \textrm{\boldmath $\pi$} } \leq \|{\bf P}\|^{\tau}_{ \textrm{\boldmath $\pi$}} \Big\| \frac{\textrm{\boldmath $\mu$}(0)}{\textrm{\boldmath $\pi$} } -1 \Big\| _{2, \textrm{\boldmath $\pi$} } \leq e_{max}^{\tau} \Big\| \frac{\textrm{\boldmath $\mu$}(0)}{\textrm{\boldmath $\pi$} } -1 \Big\| _{2, \textrm{\boldmath $\pi$} }  .
\end{equation}

Since
\begin{eqnarray}
\Big\| \frac{\textrm{\boldmath $\mu$}(0)}{\textrm{\boldmath $\pi$} } -1 \Big\| _{2, \textrm{\boldmath $\pi$} } &=& \sqrt{\sum_{i \in \textrm{\boldmath $\Omega$}} {\pi(i) \Big( \frac{\mu(i, 0)}{\pi(i)} - 1\Big)^2}} {}\nonumber \\
& \leq & \sqrt{\frac{1}{\min_{i}\pi(i)}},
\end{eqnarray}
for any $\delta > 0$, we have $\Big\| \frac{\textrm{\boldmath $\mu$}(\tau)}{\textrm{\boldmath $\pi$} } -1 \Big\| _{2, \textrm{\boldmath $\pi$} } \leq \delta$ if
\begin{equation}
\tau \geq \frac{\frac{1}{2}\log1/\pi_{min} + \log 1/\delta}{\log1/e_{max} },
\end{equation}
where $\pi_{min} = \min_i{\pi_i}$. The equation above suggests that the mixing time of a reversible Markov chain with transition matrix {\bf P} scales with $1- e_{max}$, where $e_{max} =  \max \{ |e| \neq 1: \textrm{e is an eigenvalue of {\bf P}} \}$. Therefore, in the following, we will refer to the mixing time of a reversible Markov chain with transition matrix {\bf P} as: $T_{mix} = \frac{1}{1-e_{max}}$.

Recall that following the updating rules of DISQUO algorithm, there are at most $N$ updates at every time slot, where $N$ is the number of ports. Therefore, we will consider a multiple-update Glauber dynamics defined as follows. 

\begin{definition}
(Multiple-update Glauber dynamics) Consider a graph ${\bf G(V, E)}$,  with ${\bf W} = [W_i]_{i \in {\bf V}}$, which is a vector of weights associated with the vertices. Multiple update Glauber dynamics (MUGD) is a Markov chain over $\mathcal{I}({\bf G})$. Suppose that the chain is at state ${\bf X}(n-1)= [X_i(n-1)]_{i \in {\bf V}}$ at time $n-1$. The next transition of multiple-update Glauber dynamics follows the rules:
\begin{itemize}
\item Randomly pick a set ${\bf H}(n) \in $ $\mathcal{I}(G)$ at random.
\item For $i \in {\bf H}(n)$:
	\begin{itemize}
		\item If $\forall j \in $$\mathcal{N}$$(i)$, $X_j(n-1) = 0$, then
			\begin{displaymath}
				X_i(n) = \left \{ \begin{array}{ll} 
					          1 & \textrm{with probability $\frac{\exp(W_i)}{1+\exp(W_i)}$ } \\
					          0 & \textrm{otherwise.}
					          \end{array} 
					          \right.
			\end{displaymath}
		\item Otherwise, $X_i(n) = 0$.
	\end{itemize}
\item $X_{i}(n)=X_{i}(n-1)$, for all $i \notin {\bf H}(n)$.
\end{itemize}
\end{definition}

The transition matrix is similar to Eq. (\ref{eq:tran_prob}) but with a vector of fixed weights. The multiple-update Glauber dynamics is also a positive recurrent, time-reversible Markov chain. It is easy to verify that the product-form stationary distribution in Eq. (\ref{eq:dynamics_sd}) satisfies the detailed balance equation in Eq. (\ref{eq:balance}) that it is also the stationary distribution of the multiple-update Glauber dynamics. In the following lemma, we will give an upper bound on the mixing time of the multiple-update Glauber dynamics.

\begin{lemma}
(Mixing time of multiple-update Glauber dynamics) Let {\bf P} be the transition matrix of the multiple-update Glauber dynamics on a graph ${\bf G=(V, E)}$, for which there are $N$ vertices with weights ${\bf W} = [W_i]_{i \in {\bf V}}$. We have:
\begin{equation}
T_{mix} \leq 2^{6N}\exp(4NW_{max}),
\end{equation}
where $W_{max} = \max_{i \in {\bf V}} W_i$.
\end{lemma}

\begin{IEEEproof}
For a nonempty set ${\bf A} \subset \mathcal{I}(G)$, we have:
\begin{displaymath}
\pi({\bf A}) = \sum_{i \in {\bf A}}{\pi(i)}.
\end{displaymath}
Let us define the following:
\begin{displaymath}
F({\bf A}) = \sum_{i \in {\bf A}, j \in {\bf A}^c}{\pi(i) p_{ij}}. 
\end{displaymath}
The conductance of the transition matrix ${\bf P}$ is defined as:
\begin{displaymath}
\phi({\bf P}) = \min_{{\bf A} \subset \textrm{$\mathcal{I}$}(G): \pi({\bf A}) \leq \frac{1}{2}} \frac{F({\bf A})}{\pi({\bf A})}.
\end{displaymath}
There is a well-known conductance bound \cite{cbound01, cbound02} with the form:
\begin{displaymath}
e_{max} \leq 1 - \frac{\phi^2({\bf P})}{2}.
\end{displaymath}
Now, we have:
\begin{eqnarray}
\phi({\bf P}) &=& \min_{{\bf A} \subset \textrm{$\mathcal{I}$}(G): \pi({\bf A}) \leq \frac{1}{2}} \frac{F({\bf A})}{\pi({\bf A})} {}\nonumber \\
& = & \min_{{\bf A} \subset \textrm{$\mathcal{I}$}(G): \pi({\bf A}) \leq \frac{1}{2}} \frac{ \sum_{{\bf X} \in {\bf A}, {\bf X}'  \in {\bf A}^c } {\pi({\bf X}) P({\bf X}, {\bf X}')} } {\pi({\bf A})} {}\nonumber \\
& \geq & 2\min_{{\bf A} \in \mathcal{I}(G)} {P({\bf A}, {\bf A}^c)} {} \nonumber \\
& \geq & 2\min_{P({\bf X}, {\bf X}') \neq 0} { \pi({\bf X}) P({\bf X}, {\bf X}') } {}\nonumber \\
& \geq & 2\min_{{\bf X}} { \pi({\bf X}) } \min_{{\bf X} \neq {\bf X}', P({\bf X}, {\bf X}') \neq 0}{ P({\bf X}, {\bf X}') } {}\nonumber 
\end{eqnarray}
For the Glauber dynamics, the stationary distribution can be lower bounded by:
\begin{eqnarray}
\pi({\bf X}) &\geq& \frac{1}{\sum_{{\bf X} \in \mathcal{I}(G)} {\exp(\sum_{i \in {\bf X}} {W_i})} } \nonumber \\
	& \geq & \frac{1}{ | \textrm{$\mathcal{I}$}(G)| \exp (N W_{max}) } \nonumber \\
	& \geq & \frac{1}{ 2^N \exp (N W_{max}) } \nonumber
\end{eqnarray}
Also, we have:
 \begin{displaymath}
 P({\bf X}, {\bf X}') \geq \frac{1}{2^N} \Big( \frac{1}{1+\exp(W_{max})} \Big)^N.
 \end{displaymath}
 So,
 \begin{eqnarray}
 \phi ({\bf P}) &\geq& \frac{2 }{ 2^{2N} (1+\exp(W_{max}))^N \exp(NW_{max}) } \nonumber \\
 & \geq & \frac{2}{2^{3N}\exp(2NW_{max})} \nonumber
 \end{eqnarray}
 Thus, 
 \begin{displaymath}
 e_{max} \leq 1- \frac{2}{2^{6N}\exp(4NW_{max})} \leq 1- \frac{1}{2^{6N}\exp(4NW_{max})}.
  \end{displaymath}
 
 Since $T_{mix} = \frac{1}{1-e_{max}}$, we have:
  \begin{displaymath}
T_{mix} \leq 2^{6N}\exp(4NW_{max}).
  \end{displaymath}
 
\end{IEEEproof}

Now, we are ready to prove Lemma \ref{lem:convergence}. We will first identify the condition for the system to converge in Lemma \ref{lem:chap12_cong}. Then, in Lemma \ref{lem:chap12_wf}, we will prove that if the weight function $f(\cdot)$ are well designed, the condition for the system convergence can be satisfied, and thus finish the proof of Lemma \ref{lem:convergence}. The proof of Lemma \ref{lem:chap12_cong} is mainly adapted from Ref. \cite{aloha08, aloha0811}.

Let ${\bf P}_n$ denote the transition matrix at time $n$. $e_{max}(n) = \max\{ |e|:  |e| \neq 1, \textrm{e is the eigenvalue of }{\bf P}_n\}$, and $T_{n} = \frac{1}{1-e_{max}(n)}$, which is the mixing time of the multiple-update Glauber dynamics with weight vector ${\bf W}(n)$ .

In the following Lemma, we will prove that given the condition that $\alpha_n T_{n+1} \leq {\delta}/{8}$ ($\forall \delta > 0$), the system can converge within finite time, where $\alpha_n $ is defined as Eq. (\ref{eq:alphan}). We will also give an upper bound on the mixing time of the system. 
\begin{lemma}
\label{lem:chap12_cong}
If $\alpha_n T_{n+1} \leq {\delta}/{8}$, then for any $\delta > 0$, $\| \textrm{\boldmath $\mu$}_n - \textrm{\boldmath $\pi$}_n \|_{TV} \leq \delta$ holds for all $n \geq n*$, where $T_{n+1}$ is the mixing time of the multiple-update  Glauber dynamics with weight vector ${\bf W}(n+1)$, 
\begin{equation}
\label{eq:alphan}
\alpha_n = \sum_{i,j}{f'(\tilde{Q}_{ij}(n)) + f'(\tilde{Q}_{ij}(n+1))},
\end{equation}
and 
\begin{equation}
n^* = \min_{n} \sum_{i=1}^{n}{{\frac{1}{T_i^2}} \geq \log(\frac{2}{\delta}) + \frac{N^2}{2}\Big( \log2 + W_{max}(0) \Big). },
\end{equation}

\end{lemma}

\begin{IEEEproof}
The stationary distributions for the multiple-update Glauber dynamics with weight vectors ${\bf W}(n)$ and ${\bf W}(n+1)$ can be written as:
\begin{displaymath}
\textrm{\boldmath $\pi$}_n ({\bf X}) = \frac{1}{Z_n} \exp(\sum_{(i, j) \in {\bf X}} { W_{ij}(n) }),
\end{displaymath}
and 
\begin{displaymath}
\textrm{\boldmath $\pi$}_{n+1}  ({\bf X}) = \frac{1}{Z_{n+1}} \exp(\sum_{(i, j) \in {\bf X}} { W_{ij}(n+1) }),
\end{displaymath}
respectively. So,
\begin{equation}
\label{eq:chap12_a}
\frac{ \textrm{\boldmath $\pi$}_{n+1} ({\bf X})  }{ \textrm{\boldmath $\pi$}_{n} ({\bf X})  } = \frac{Z_{n}}{Z_{n+1}} \exp \Big( \sum_{(i, j) \in {\bf X}} { (W_{ij}(n+1) -W_{ij}(n)) } \Big),
\end{equation}
and
\begin{eqnarray}
\frac{Z_{n}}{Z_{n+1}} &\leq& \frac{ \sum_{{\bf X} \in \textrm{$\mathcal{I}$}({\bf G})} { \exp ( \sum_{ (i, j) \in {\bf X} } { W_{ij}(n) } ) } }{ \sum_{{\bf X} \in \textrm{$\mathcal{I}$}({\bf G})} { \exp ( \sum_{ (i, j) \in {\bf X} } { W_{ij}(n+1) } ) } } \nonumber \\
&\leq& \max_{{\bf X}} {  \exp ( \sum_{ (i, j) \in {\bf X} } { W_{ij}(n) - W_{ij}(n+1) } ) }. \nonumber
\end{eqnarray}
Note that $W_{ij}(n) = f(\tilde{Q}_{ij}(n))$, and $f(\cdot)$ is a increasing concave function that $f(b) - f(a) \leq f'(a)(b-a)$. Therefore, 
\begin{eqnarray}
W_{ij}(n) - W_{ij}(n+1) &\leq& f'(\tilde{Q}_{ij}(n+1))(\tilde{Q}_{ij}(n) - \tilde{Q}_{ij}(n+1)) \nonumber \\
& \leq & f'(\tilde{Q}_{ij}(n+1)). \nonumber
\end{eqnarray}
The equation above is according to the fact that at every time slot, there is at most one arrival and one departure that we have $-1 \leq \tilde{Q}_{ij}(n) - \tilde{Q}_{ij}(n+1) \leq 1$. So,
\begin{eqnarray}
\label{eq:chap12_b}
\frac{Z_{n}}{Z_{n+1}} &\leq& \max_{{\bf X}} {  \exp ( \sum_{(i, j) \in {\bf X}}{f'(\tilde{Q}_{ij}(n+1))} ) } \nonumber \\
&\leq& \exp ( \sum_{i, j}{f'(\tilde{Q}_{ij}(n+1))} )
\end{eqnarray}
Similarly, we have
\begin{equation}
\frac{Z_{n+1}}{Z_n}  \leq \exp ( \sum_{i, j}{f'(\tilde{Q}_{ij}(n))} )
\end{equation}

From Eq. (\ref{eq:chap12_a}) and Eq. (\ref{eq:chap12_b}), we have:
\begin{eqnarray}
\frac{ \textrm{\boldmath $\pi$}_{n+1} ({\bf X})  }{ \textrm{\boldmath $\pi$}_{n} ({\bf X})  }  & = &\frac{Z_{n}}{Z_{n+1}} \exp \Big( \sum_{(i, j) \in {\bf X}} { (W_{ij}(n+1) -W_{ij}(n)) } \Big) \nonumber \\
&\leq& \exp(\sum_{i, j} {f'(\tilde{Q}_{ij}(n)) + f'(\tilde{Q}_{ij}(n+1))  }), 
\end{eqnarray}
and also, 
\begin{equation}
\frac{ \textrm{\boldmath $\pi$}_{n} ({\bf X})  }{ \textrm{\boldmath $\pi$}_{n+1} ({\bf X})  }  \leq \exp(\sum_{i, j} {f'(\tilde{Q}_{ij}(n)) + f'(\tilde{Q}_{ij}(n+1))  }). 
\end{equation}
Define $\alpha_n = \sum_{i,j}{ \Big[ f'(\tilde{Q}_{ij}(n)) + f'(\tilde{Q}_{ij}(n+1)) \Big]}$. We have:
\begin{equation}
\exp(-\alpha_n) \leq \frac{ \textrm{\boldmath $\pi$}_{n} ({\bf X})  }{ \textrm{\boldmath $\pi$}_{n+1} ({\bf X})  }   \leq \exp(\alpha_n)
\end{equation}
Recall that $\alpha_n T_{n+1} \leq \frac{\delta}{8}$, and $T_{n+1}= \frac{1}{1 - e_{max}(n+1)} \geq 1$ is the mixing time of the multiple-update  Glauber dynamics with weight vector ${\bf W}(n)$. Since $\delta$ is any small positive number, we have $0< \alpha_n < 1$. Since $1-x \leq e^{-x}$ and $e^x \leq 1+2x$ for all $x \in [0, 1]$, we have
\begin{displaymath}
-\alpha_n \leq \frac{ \textrm{\boldmath $\pi$}_{n} ({\bf X})  }{ \textrm{\boldmath $\pi$}_{n+1} ({\bf X})  }  -1 \leq 2\alpha_n.
\end{displaymath}
So, 
\begin{displaymath}
\Big( \frac{ \textrm{\boldmath $\pi$}_{n} ({\bf X})  } { \textrm{\boldmath $\pi$}_{n+1} ({\bf X})  }  -1 \Big)^2 \leq 4\alpha_n^2.
\end{displaymath}
Then, 
\begin{eqnarray}
\| \textrm{\boldmath $\pi$}_{n+1} - \textrm{\boldmath $\pi$}_{n} \|^2_{2, 1/\textrm{\boldmath $\pi$}_{n+1} } &=& \| \frac{\textrm{\boldmath $\pi$}_{n}}{\textrm{\boldmath $\pi$}_{n+1}} -1 \|^2_{2, \textrm{\boldmath $\pi$}_{n+1} } \nonumber \\
& = & \sum_{{\bf X}} { \pi_{n+1}({\bf X}) \Big( \frac{\pi_n({\bf X})}{\pi_{n+1}({\bf X})} - 1 \Big)^2 } \nonumber \\
& \leq & 4\alpha^2_n \sum_{{\bf X}} { \pi_{n+1}({\bf X}) } =  4\alpha^2_n \nonumber 
\end{eqnarray}
Thus, 
\begin{displaymath}
\| \textrm{\boldmath $\pi$}_{n+1} - \textrm{\boldmath $\pi$}_{n} \|_{2, 1/\textrm{\boldmath $\pi$}_{n+1} } \leq 2\alpha_n.
\end{displaymath}
The distance between $\textrm{\boldmath $\mu$}_n$ and $\textrm{\boldmath $\pi$}_n$ can then be bounded by:
\begin{eqnarray}
\label{eq:chap12_c}
\| \frac{\textrm{\boldmath $\mu$}_n} {\textrm{\boldmath $\pi$}_n}  - 1 \|_{2, \textrm{\boldmath $\pi$}_n} & = & \| \textrm{\boldmath $\mu$}_n - \textrm{\boldmath $\pi$}_n \|_{2, 1/\textrm{\boldmath $\pi$}_n} \nonumber \\
&\leq & \| \textrm{\boldmath $\mu$}_n - \textrm{\boldmath $\pi$}_{n-1} \|_{2, 1/\textrm{\boldmath $\pi$}_n}  \nonumber \\
&& + \| \textrm{\boldmath $\pi$}_{n-1} - \textrm{\boldmath $\pi$}_n \|_{2, 1/\textrm{\boldmath $\pi$}_n} \nonumber \\
& \leq & \| \textrm{\boldmath $\mu$}_n - \textrm{\boldmath $\pi$}_{n-1} \|_{2, 1/\textrm{\boldmath $\pi$}_n}  + 2\alpha_{n-1}.
\end{eqnarray}
Note that 
\begin{eqnarray}
\label{eq:chap12_d}
\| \textrm{\boldmath $\mu$}_n - \textrm{\boldmath $\pi$}_{n-1} \|^2_{2, 1/\textrm{\boldmath $\pi$}_n} & = & \sum_{{\bf X}} { \frac{1}{\pi_n({\bf X})} (\mu_n({\bf X}) - \pi_{n-1}({\bf X}) )^2 } \nonumber \\
& = & \sum_{{\bf X}}  { \frac{ \pi_{n-1} ({\bf X}) } { \pi_n ({\bf X}) }  \frac{1} { \pi_{n-1} ({\bf X}) } } \nonumber \\
&& \cdot {  ( \mu_n({\bf X}) - \pi_{n-1}({\bf X}) )^2 }  \nonumber \\
&\leq& e^{(\alpha_{n-1})} \| \textrm{\boldmath $\mu$}_n - \textrm{\boldmath $\pi$}_{n-1} \|^2_{2, 1/\textrm{\boldmath $\pi$}_{n-1} }.
\end{eqnarray}
From Eq. (\ref{eq:chap12_c}) and (\ref{eq:chap12_d}), we have:
\begin{eqnarray}
\label{eq:chap12_f}
\| \frac{\textrm{\boldmath $\mu$}_n} {\textrm{\boldmath $\pi$}_n}  - 1 \|_{2, \textrm{\boldmath $\pi$}_n} &\leq& \exp(\alpha_{n-1}/2) \| \textrm{\boldmath $\mu$}_n - \textrm{\boldmath $\pi$}_{n-1} \|_{2, 1/\textrm{\boldmath $\pi$}_{n-1} } \nonumber \\
&& + 2\alpha_{n-1} \nonumber \\
&\leq& (1+\alpha_{n-1}) \| \textrm{\boldmath $\mu$}_n - \textrm{\boldmath $\pi$}_{n-1} \|_{2, 1/\textrm{\boldmath $\pi$}_{n-1} } \nonumber \\
&& + 2\alpha_{n-1}
\end{eqnarray}
Let us define 
\begin{equation}
\beta_n =  \| \textrm{\boldmath $\mu$}_{n+1} - \textrm{\boldmath $\pi$}_{n} \|_{2, 1/\textrm{\boldmath $\pi$}_{n} }.
\end{equation}
Note that $\alpha_n \leq \alpha_n T_{n+1} \leq \delta/8$. If $\beta_n \leq \delta/2$, then from Eq. (\ref{eq:chap12_f}), we have:
\begin{equation}
\| \frac{\textrm{\boldmath $\mu$}_n} {\textrm{\boldmath $\pi$}_n}  - 1 \|_{2, \textrm{\boldmath $\pi$}_n} \leq \delta, 
\end{equation}
for all $n > n^*$. Therefore, to establish the result, we then have to prove that $\beta_n \leq \delta/2$ holds for any $n > n^*$. Consider the following equation:
\begin{eqnarray}
\label{eq:chap12_e}
\beta_{n+1} & =  &\| \textrm{\boldmath $\mu$}_{n+2} - \textrm{\boldmath $\pi$}_{n+1} \|_{2, 1/\textrm{\boldmath $\pi$}_{n+1} } \nonumber \\
& =  &\| \frac{ \textrm{\boldmath $\mu$}_{n+2} }{ \textrm{\boldmath $\pi$}_{n+1} } - 1\|_{2, \textrm{\boldmath $\pi$}_{n+1} } \nonumber \\
& = &\| \frac{ \textrm{\boldmath $\mu$}_{n+1} {\bf P}_{n+1} }{ \textrm{\boldmath $\pi$}_{n+1} } - 1 \|_{2, \textrm{\boldmath $\pi$}_{n+1} } \nonumber \\
& \leq & \| {\bf P}_{n+1} \|_{\textrm{\boldmath $\pi$}_{n+1} } \| \textrm{\boldmath $\mu$}_{n+1}  - \textrm{\boldmath $\pi$}_{n+1} \|_{2, 1/\textrm{\boldmath $\pi$}_{n+1} } \nonumber \\
& \leq & e_{max}(n+1)  \| \textrm{\boldmath $\mu$}_{n+1}  - \textrm{\boldmath $\pi$}_{n+1} \|_{2, 1/\textrm{\boldmath $\pi$}_{n+1} }  \nonumber \\
& \leq & (1-\frac{1}{T_{n+1}}) \nonumber \\
&& \cdot \Big( \| \textrm{\boldmath $\mu$}_{n+1}   - \textrm{\boldmath $\pi$}_{n} \|_{2, \frac{1}{\textrm{\boldmath $\pi$}_{n+1} } } +  \| \textrm{\boldmath $\pi$}_{n}    - \textrm{\boldmath $\pi$}_{n+1} \|_{2, \frac{1}{\textrm{\boldmath $\pi$}_{n+1} } } \Big) \nonumber \\
& \leq & (1-\frac{1}{T_{n+1}})\Big( \| \textrm{\boldmath $\mu$}_{n+1}   - \textrm{\boldmath $\pi$}_{n} \|_{2, 1/\textrm{\boldmath $\pi$}_{n+1} }  + 2\alpha_n \Big) \nonumber \\
&\leq &(1-\frac{1}{T_{n+1}}) \nonumber \\
&& \cdot \Big( \exp(\alpha_n/2) \| \textrm{\boldmath $\mu$}_{n+1}   - \textrm{\boldmath $\pi$}_{n} \|_{2, 1/\textrm{\boldmath $\pi$}_{n} }  + 2\alpha_n \Big) \nonumber \\
&= &(1-\frac{1}{T_{n+1}}) \Big( \exp(\alpha_n/2) \beta_n  + 2\alpha_n \Big) \nonumber \\
&\leq&(1-\frac{1}{T_{n+1}}) \Big( (1+\alpha_n)\beta_n  + 2\alpha_n \Big) 
\end{eqnarray}
Suppose that $\beta_n \leq \delta /2$, Eq. (\ref{eq:chap12_e}) can be written as:
\begin{eqnarray}
\label{eq:chap12_g}
\beta_{n+1} &\leq & (1-\frac{1}{T_{n+1}}) \Big( \frac{\delta}{2}  + (2+\frac{\delta}{2})\alpha_n \Big) \nonumber \\
&\leq& (1-\frac{1}{T_{n+1}}) \Big( \frac{\delta}{2}  + (2+\frac{\delta}{2}) \frac{\delta}{8T_{n+1}} \Big) \nonumber \\
&\leq& \frac{\delta}{2} - \frac{1}{T_{n+1}} \Big( \frac{\delta}{2}  + (2+\frac{\delta}{2}) \frac{\delta}{8T_{n+1}} - (2+\frac{\delta}{2}) \frac{\delta}{8} \Big) \nonumber \\
&\leq&\frac{\delta}{2}.
\end{eqnarray}
From Eq. (\ref{eq:chap12_g}), we can see that, if $\beta_{n^*} \leq \delta/2$, then for any $n > n^*$,  $\beta_n \leq \delta/2$. 

In the following, we will find $n^*$ which is the smallest number to satisfy $\beta_{n} \leq \delta/2$. Note that if $\beta_{n^*} \leq \delta/2$, then for any $n > n^*$, $\beta_n \leq \delta/2$. Therefore, for any $n < n^*$, $\beta_n > \delta/2$. So, for $n < n^*$, we have
\begin{eqnarray}
\beta_n &\leq& (1-\frac{1}{T_n}) \Big(  (1+\alpha_{n-1})\beta_{n-1} + 2 \alpha_{n-1} \Big) \nonumber \\
&\leq& (1-\frac{1}{T_n}) \Big(  (1+\alpha_{n-1})\beta_{n-1} + 4 \alpha_{n-1} \frac{\beta_{n-1}}{\delta} \Big) \nonumber \\
&\leq& (1-\frac{1}{T_n}) \Big(  1+ (1+  \frac{4}{\delta})\alpha_{n-1} \Big) \beta_{n-1} \nonumber \\
&\leq& (1-\frac{1}{T_n}) \Big(  1+ \frac{1}{T_{n}} \Big) \beta_{n-1} \nonumber \\
& = & (1 - \frac{1}{T_n^2}) \beta_{n-1}\nonumber \\
&\leq & \exp(- \frac{1}{T_n^2})\beta_{n-1}\nonumber \\
&\leq& \exp(-\sum_{i=1}^n{\frac{1}{T_i^2}}) \beta_0,
\end{eqnarray}
where $\beta_0$ can be written as:
\begin{eqnarray}
\beta_0 & = &  \| \frac{ \textrm{\boldmath $\mu$}_1 }{ \textrm{\boldmath $\pi$}_0 } - 1\|_{2, \textrm{\boldmath $\pi$}_0} \nonumber \\
& = &  \| \textrm{\boldmath $\mu$}_0 {\bf P}_0 - \textrm{\boldmath $\pi$}_0 \|_{2, 1/\textrm{\boldmath $\pi$}_0} \nonumber \\
& = &  e_{max}(0) \| \textrm{\boldmath $\mu$}_0  - \textrm{\boldmath $\pi$}_0 \|_{2, 1/\textrm{\boldmath $\pi$}_0} \nonumber \\
& \leq &\sqrt{ \frac{1}{\pi_{min}(0)} },
\end{eqnarray}
where $\pi_{min}(0) = \min_i{\pi_0(i)} \geq \frac{1}{Z(0)} > \frac{1}{2^{N^2} \exp(N^2 W_{max}(0))}$. So,
\begin{equation}
\beta_0 \leq \Big( 2\exp(W_{max}(0)) \Big)^{N^2/2}.
\end{equation}
$\beta_{n^*} \leq \delta /2$ that it satisfies the condition:
\begin{equation}
\Big( 2\exp(W_{max}(0)) \Big)^{N^2/2} \exp(-\sum_{i=1}^{n^*}{\frac{1}{T_i^2}}) \leq \delta/2, 
\end{equation}
or,
\begin{equation}
 \sum_{i=1}^{n^*}{\frac{1}{T_i^2}} \geq \log(\frac{2}{\delta}) + \frac{N^2}{2}\Big( \log2 + W_{max}(0) \Big). 
\end{equation}
Note that $T_i$ is bounded such that there always exists a $n^*$ which can satisfies the condition above. 
\end{IEEEproof}

\begin{lemma}
\label{lem:chap12_wf}
If $f(x) = \frac{\log(1+x)}{g(x)}$, there exists a constant C that when $\| {\bf Q} \| > C$, for any $\delta > 0$, $\alpha_n T_{n+1} \leq \delta/8$, where $g(x)$ is a function that satisfies the following conditions:
\begin{itemize}
	\item $g(x) \geq 1$, for all $x \geq 0$.
	\item $g'(x) \geq 0$, for all $x \geq 0$.
	\item $\lim_{x \to \infty } g(f^{-1}(x)) = \infty$.
\end{itemize} 
\end{lemma}

\begin{IEEEproof}
We have:
\begin{equation}
f'(x) = \frac{1}{(1+x)g(x)} - \frac{\log(1+x) g'(x)}{g^2(x)}\leq \frac{1}{1+x}.
\end{equation}
Also,
\begin{equation}
f^{-1}(x) = \exp(xg(f^{-1}(x))) -1.
\end{equation}
Recall that
\begin{eqnarray} 
\alpha_n &=& \sum_{i,j}{f'(\tilde{Q}_{ij}(n)) + f'(\tilde{Q}_{ij}(n+1))} \nonumber \\
&\leq &N^2 ( f'(\tilde{Q}_{min}(n) + f'(\tilde{Q}_{min}(n+1)), 
\end{eqnarray}
where $\tilde{Q}_{min}(n) = \min_{ij}{\tilde{Q}_{ij}(n)}$, and 
\begin{equation}
T_{n+1} \leq 2^{6N^2}\exp(4N^2W_{max}(n+1))
\end{equation}
So,
\begin{eqnarray}
\alpha_n T_{n+1} &\leq& 2^{6N^2}\exp(4N^2W_{max}(n+1))  \nonumber \\
&& \cdot N^2 ( f'(\tilde{Q}_{min}(n) + f'(\tilde{Q}_{min}(n+1)) \nonumber \\
&\leq& N^2 2^{6N^2}\exp(4N^2W_{max}(n+1)) \nonumber \\
&& \cdot  ( \frac{1}{1+\tilde{Q}_{min}(n)} + \frac{1}{1+\tilde{Q}_{min}(n+1)}) \nonumber \\
&\leq& { 2N^2 2^{6N^2}\exp(4N^2W_{max}(n+1)) } \nonumber \\
&& \cdot  ( \frac{1}{\tilde{Q}_{min}(n+1)}) \nonumber \\
&\leq& { 2N^2 2^{6N^2} \exp \Big( 4N^2 (  \frac{2N^2}{ \epsilon} ) W_{min}(n+1) \Big)} \nonumber \\
&& \cdot  \frac{1}{f^{-1}(W_{min}(n+1)) } \nonumber \\
&=& {2 N^2 2^{6N^2}\exp \Big( 8N^4 / \epsilon W_{min}(n+1) \Big)} \nonumber \\
& &\cdot   \frac{1}{ \exp \Big(W_{min}(n+1)g(f^{-1}(W_{min}(n+1))) \Big) - 1 }.\nonumber  
\end{eqnarray}
Then,
\begin{eqnarray}
\alpha_n T_{n+1} &\leq & \exp \Big[ \Big( 8N^4 / \epsilon \nonumber \\
 && \textrm{          } - g(f^{-1}(W_{min}(n+1)))  \Big) W_{min}(n+1) \Big] \nonumber \\
& & \cdot   2N^2 2^{6N^2} \Big(1+ \frac{1}{f^{-1}(W_{min}(n+1)) } \Big)
\end{eqnarray}
If $W_{max} \to \infty$, $W_{min} \to \infty$ such that $g(f^{-1}(W_{min}(n+1))) \to \infty$, and thus the value of $8N^4 / \epsilon - g(f^{-1}(W_{min}(n+1))) \to \infty$. Therefore, for any $\delta > 0$, there exists a constant $C$ such that when $\| {\bf Q}\| \geq C$,  $\alpha_n T_{n+1} \leq \delta/8$ holds. 

By proving Lemma \ref{lem:chap12_wf}, we give the sufficient condition that the system can converge, as stated in Lemma \ref{lem:chap12_cong}. Therefore, following the randomized scheduling algorithm, the inhomogeneous Markov chain can still converge to a stationary distribution, which can be expressed as Eq. (\ref{eq:sd}). We finish the proof of Lemma \ref{lem:convergence}. 
\end{IEEEproof}

\end{document}